\documentclass[pra,twocolumn,showpacs,preprintnumbers,amsmath,amssymb,superscriptaddress]{revtex4}
\usepackage{bm, psfrag, graphics}
\begin{document}

\title{Multiconfigurational Hartree-Fock theory for identical bosons in a double well}
\author{D. Masiello}
\email{masiello@u.washington.edu}
\affiliation{Department of Chemistry, University of Washington, Seattle, Washington 98195-1700, USA}
\author{S. B. McKagan}
\altaffiliation[Present address: ]{JILA, University of Colorado, Boulder, Colorado 80309-0440, USA}
\affiliation{Department of Physics, University of Washington, Seattle, Washington 98195-1560, USA}
\author{W. P. Reinhardt}
\affiliation{Department of Chemistry, University of Washington, Seattle, Washington 98195-1700, USA}
\affiliation{Department of Physics, University of Washington, Seattle, Washington 98195-1560, USA}
\date{\today}       
\begin{abstract}
Multiconfigurational Hartree-Fock theory is presented and implemented in an investigation of the fragmentation of a Bose-Einstein condensate made of identical bosonic atoms in a double well potential at zero temperature. The approach builds in the effects of the condensate mean field and of atomic correlations by describing generalized many-body states that are composed of multiple configurations which incorporate atomic interactions. Nonlinear and linear optimization is utilized in conjunction with the variational and Hylleraas-Undheim theorems to find the optimal ground and excited states of the interacting system. The resulting energy spectrum and associated eigenstates are presented as a function of double well barrier height. Delocalized and localized single configurational states are found in the extreme limits of the simple and fragmented condensate ground states, while multiconfigurational states and macroscopic quantum superposition states are revealed throughout the full extent of barrier heights. Comparison is made to existing theories that either neglect mean field or correlation effects and it is found that contributions from both interactions are essential in order to obtain a robust microscopic understanding of the condensate's atomic structure throughout the fragmentation process.
\end{abstract}
\maketitle

\section{Introduction}
Recent experimental realization of a trapped atom interferometer using a coherently split Bose-Einstein condensate (BEC) \cite{Shin2004a,Saba05} as well as the direct observation of tunneling and self-trapping in weakly linked BECs \cite{Anker04,Albiez04} have provided an impetus to formulate a comprehensive theoretical description of the zero temperature Bose gas confined to a trapping potential that can be continuously deformed from a single well into a double well with large barrier height. Such a deformation of the trapping potential impels the BEC from a single coherent entity to a fragmented condensate made up of two coherent and potentially correlated moieties. The quantum many-body physics governing this complex fragmentation process involves competition and balance between the effects of the condensate's mean field on the interacting atomic gas and the correlations that emerge between atoms in different Fock states.

Various theoretical descriptions of the simple and fragmented BEC exist in the literature today. At both low and high barrier heights, the many-body ground state of the condensate is well approximated in mean field theory by a single Fock state, which expresses a particular arrangement of atoms among one, two, or even many single-particle states \cite{Cederbaum2003a,Cederb04,Alon05}. If only one single-particle state is involved, then mean field theory reduces to the Hartree theory \cite{Hartree1928a}, which further reduces to the standard Gross-Pitaevskii formalism \cite{Gross1961a,Pitaevskii1961a} upon invoking the contact interaction approximation \cite{Huang1957a,Huang1987a}. More elaborate mean field theories, such as Hartree-Fock \cite{Faddeev}, are built by utilizing more than one single-particle state and imposing the correct symmetrization due to the indistinguishability of identical particles. Where mean field equations specify the underlying single-particle states, we call the single Fock state a single {\it configuration} \footnote{We choose to use this terminology, which is adapted from quantum chemistry, because we wish to emphasize the connection between the atomic structure theory of cold atomic gases and the well established theory of electronic structure in atoms and molecules. This relationship has already been promoted in the BEC literature \cite{Esry1997b}.}. However, more complex many-body states exist, at all barrier heights, that are made up of superpositions of multiple configurations.

As mean field theory describes only a single configuration, it necessarily lacks all correlation effects that arise between configurations. To this end, multiconfigurational approaches have been attempted, which when applied to cold atomic gases in a double well trap geometry, are largely based on two-well limits of continuum lattice models such as the Bose-Hubbard model. Both atomic correlations and, to a partial extent, mean field interactions are included automatically in these theories. Correlations emerge between configurations while mean field interactions occur in two places: first, directly through the interaction term in the many-body Hamiltonian, and second in the underlying single-particle states that make up the matrix elements in the Hamiltonian as well as each configuration. Previous multiconfigurational efforts found in the BEC literature include only the first type of mean field effect and therefore do not describe general many-body states of the system by superpositions of configurations but only those where the effects of atom-atom interactions on the shape of the single-particle states are neglected. In these works, the underlying single-particle states have been chosen either to be solutions of the single-particle Schr\"odinger equation \cite{Spekkens1999a} or parameters have been introduced to replace the matrix elements in the Hamiltonian altogether \cite{Milburn1997a,Steel1998a,Raghavan1999a,Reinhardt2003a,Mahmud2003a}. For example, in the Bose-Hubbard model $U$ is a site energy and $J$ is a tunneling parameter, and neither depends on $N,$ the particle number, and neither is computed from first principles. In other words, mean field effects are not fully included in these treatments: the single-configurational and multiple Fock state approaches are separate and complementary.

A more complete description should characterize a general state of the system by a superposition of many configurations in which the underlying single-particle states include the effects of the condensate mean field. Such a theory would be capable of illuminating the quantum many-body structure of the BEC {\it throughout} the fragmentation process as well as at the extreme simple and fragmented limits. It is the purpose of the present paper to provide such a formulation.

The challenge to formulate such a theory has already been partially fulfilled by several authors \cite{Cederbaum2003a,Cederb04,Alon05,Spekkens1999a,Milburn1997a,Steel1998a,Raghavan1999a,Reinhardt2003a,Mahmud2003a}. However, to our knowledge, no satisfactory work has been presented in the cold atom context that fully addresses this task. To this end, we formulate a new approach that variationally combines the Hartree-Fock mean field theory for $N$ identical bosons in two single-particle states with a full diagonalization of the many-body Hamiltonian restricted to a basis of $N+1$ generalized configurations stemming from each Hartree-Fock configuration. This allows for a description of the Bose gas at zero temperature where atomic correlations emerge between configurations into which mean field effects are built. Due to its composition, this approach incorporates both types of interaction and serves as a didactic device for elucidating where they become important in the fragmentation process as well as how these two intertwined but distinct effects change the system. By choosing the atomic interaction strength to be zero in the bosonic Hartree-Fock mean field equations, our approach recovers the Schr\"odinger based model developed in \cite{Spekkens1999a}, while limiting our generalized many-body state to a single configuration recovers the mean field theories of \cite{Cederbaum2003a,Cederb04}. A preliminary investigation of our formalism can be found in \cite{McKagan04}. We acknowledge that a fermionic analog of our model, called multiconfigurational self-consistent field theory, is widely known in quantum chemistry and has been quite successful in accurately describing atomic and molecular electronic structure and geometry both at equilibrium and at dissociation \cite{Shepard1987a,Levine,Ostlund}. In fact, a proper description of dissociation of polyatomic molecules in the Born-Oppenheimer approximation \cite{Shepard1987a,Levine,Ostlund} is closely related to the problem of fragmentation of a BEC into two or more fragments. Related time-independent \cite{Bowman86,Bowman03} and time-dependent \cite{Meyer00} multiconfigurational approaches have also been developed to treat molecular vibrations at the Hartree level. In the spirit of these efforts, we refer to the work developed in this paper as the {\it multiconfigurational bosonic Hartree-Fock theory.}

To establish a consistent and general enough notation, which can be extended to our multiconfigurational Hartree-Fock approach, we organize the paper as follows. In Section II, we review the many-body theory of a gas of $N$ identical bosons restricted to a finite model space. Ground and excited eigenstates of the system are expanded onto a basis consisting of $N+1$ Fock states made up of two single-particle states. Model calculations, which lack the effects of atomic interactions on the underlying single-particle states, are discussed where the trapping potential is deformed from an initially single well to a double well geometry. Section III is devoted to a survey of Hartree-Fock mean field theory for $N$ identical bosons in two single-particle states. It is emphasized that mean field theory describes only a single configurational state and therefore lacks the correlations described by superpositions of multiple configurations. Imaginary time integration is briefly discussed for efficient solution of the resulting coupled nonlinear differential equations. While imaginary time integration schemes are a standard method of solution for the bosonic mean field equations with one single-particle state, {\it i.e.}, for the Gross-Pitaevskii equation \cite{Dalfovo96}, or with multicomponent spinors \cite{Esry1997a}, we highlight our method of maintaining spatial orthogonality between two single-particle states. Solutions of the mean field and Schr\"odinger equations are compared at various barrier heights in the strongly interacting limit.

In the Section IV, we introduce the multiconfigurational bosonic Hartree-Fock theory. For each individual energy level, multiconfigurational Hartree-Fock states of the system are constructed from the variationally optimal linear combination of generalized configurations in which the underlying mean field states are chosen so that the partitioning of atoms between its two states allows the corresponding energy to be minimized. Lastly in Section V, a systematic investigation of the atomic structure of the BEC, as a function of barrier height, throughout the fragmentation process is carried out. This is followed in Section VI by a summary and an indication of further work needed to describe ongoing experiments. An Appendix is devoted to an informal statement of the Hylleraas-Undheim theorem, which justifies such a state-by-state use of the variational theorem. In particular, we discuss our application of this theorem to the optimization of BEC excited states and their corresponding energies.

\section{Review of Many-Boson Theory in a Restricted Basis}
The many-body Hamiltonian for a gas of $N$ identical spinless bosonic atoms of mass $m$ at zero temperature interacting via a two-body potential $V({\bf x},{\bf x}')\equiv V(|{\bf x}-{\bf x}'|)$ is given by \cite{FW}
\begin{equation}
\begin{split}
\label{hamfockpseudopotential}
\hat{H}&={\textstyle\int}\hat{\Psi}^\dagger({\bf x})\left\{({-\hbar^2}/{2m})\nabla^2+V_{\textrm{ext}}({\bf x})\right\}\hat{\Psi}({\bf x})d^{3}x\\
&\ \ \ +({1}/{2}){\textstyle\int}\hat{\Psi}^\dagger({\bf x})\hat{\Psi}^\dagger({\bf x}')V({\bf x},{\bf x}')\hat{\Psi}({\bf x}')\hat{\Psi}({\bf x})d^{3}xd^{3}x'.
\end{split}
\end{equation}
Here $\hat{\Psi}$ and $\hat{\Psi}^\dagger$ are boson field operators which satisfy the equal time commutation relations $[\hat{\Psi}({\bf x}),\hat{\Psi}^\dagger({\bf x}')]=\delta({\bf x}-{\bf x}')$ and $[\hat{\Psi}({\bf x}),\hat{\Psi}({\bf x}')]=[\hat{\Psi}^\dagger({\bf x}),\hat{\Psi}^\dagger({\bf x}')]=0.$  The external trapping potential $V_{\textrm{ext}}$ that we have in mind throughout this paper is one that can be continuously deformed from a single well to a double well geometry.

\subsection{Restriction to Two Single-Particle States}
The atomic structure of a BEC confined in a double well trapping potential at zero temperature can be reasonably described in a basis of restricted Fock states of the form \cite{Steel1998a,Spekkens1999a}
\begin{equation}
\label{2fs}
|N_1,N_2\rangle=(\hat b_1^\dagger)^{N_1}(\hat b_2^\dagger)^{N_2}|{\textrm{vac}}\rangle/\sqrt{N_1!N_2!},
\end{equation}
where $N_k$ atoms are in each of the single-particle states $|\chi_k\rangle=\hat{b}^\dagger_k|{\textrm{vac}}\rangle$ for $k=1,2$ and the total number of atoms is fixed at $N=N_1+N_2.$ Here, $|{\textrm{vac}}\rangle$ is the vacuum state in which no atoms are present. The operators $\hat{b}^\dagger_k$ and $\hat{b}_k$ are boson creation and annihilation operators that add and remove single atoms in the $|\chi_k\rangle.$ They satisfy the basic commutation relations $[\hat{b}_k,\hat{b}^\dagger_l]=\delta_{kl}$ and $[\hat{b}_k,\hat{b}_l]=[\hat{b}^\dagger_k,\hat{b}^\dagger_l]=0$ for $k,l=1,2.$ While (\ref{2fs}) is certainly not the most general eigenstate imaginable, we are interested mainly in the zero temperature properties of the condensate as its constituent atoms are exchanged between two macroscopically occupied single-particle states. Where the effects of finite temperature and of fragmentation into more than two condensates are negligible, the states (\ref{2fs}) provide a rich basis in which to explore the many-body physics of the BEC fragmentation process.

The set of all Fock states of the form (\ref{2fs}), with all possible numbers of atoms in each of the two single-particle states, exhausts the restricted $N$-boson Fock space. That is, the model space is spanned by the collection
\begin{equation}
\label{set}
\{|N,0\rangle,\ |N-1,1\rangle,\ |N-2,2\rangle,\ \ldots,\ |0,N\rangle\},
\end{equation}
which is taken as a complete set having $N+1$ elements. Therefore, it is possible to expand an eigenstate of the many-body Hamiltonian (\ref{hamfockpseudopotential}) as a linear combination of $N+1$ individual Fock states according to \cite{Steel1998a,Spekkens1999a}
\begin{equation}
\label{twomodestatesnl}
|\Psi^N\rangle_\nu=\sum_{N_1=0}^{N}C^\nu_{N_1}|N_1,N_2=N-N_1\rangle,
\end{equation}
where the expansion coefficient $C^\nu_{N_1}$ expresses the probability amplitude for the $\nu$th excited state of the system to be in $|N_1,N_2\rangle.$

The many-body Hamiltonian that is associated with this model space may be derived from (\ref{hamfockpseudopotential}) by substitution of the two-state expansion of the boson field operator 
\begin{equation}
\label{fieldoptwo}
\hat{\Psi}({\bf x})=\chi_1({\bf x})\hat{b}_1+\chi_2({\bf x})\hat{b}_2,
\end{equation}
where the expansion coefficients $\chi_k({\bf x})=\langle{\bf x}|\chi_k\rangle$ are coordinate space single-particle wavefunctions that are, as yet, unspecified. After some standard algebra, one obtains \cite{FW}
\begin{widetext}
\begin{equation}
\begin{split}
\label{hamtwomode}
\hat{H}&=h_{11}\hat{N}_1+h_{22}\hat{N}_2+(1/2)[V_{1212}+V_{1221}]\hat{N}_1\hat{N}_2+(1/2)[V_{2121}+V_{2112}]\hat{N}_2\hat{N}_1+[h_{12}+V_{1112}(\hat{N}_1-1)+V_{2221}\hat{N}_2]\hat{b}_1^\dagger\hat{b}_2\\
&\ \ \ +[h_{21}+V_{1112}\hat{N}_1+V_{2221}(\hat{N}_2-1)]\hat{b}_2^\dagger \hat{b}_1+(1/2)[V_{1111}(\hat{N}_1^2-\hat{N}_1)+V_{2222}(\hat{N}_2^2-\hat{N}_2)+V_{1122}(\hat{b}_1^\dagger\hat{b}_1^\dagger\hat{b}_2 \hat{b}_2+\hat{b}_2^\dagger\hat{b}_2^\dagger\hat{b}_1\hat{b}_1)]\\
\end{split}
\end{equation}
\end{widetext}
where $\hat{N}_k=\hat{b}_k^\dagger\hat{b}_k$ is the occupation number operator for each state $|\chi_k\rangle,$ $\hat{N}=\hat{N}_1+\hat{N}_2$ is the total particle number operator, and
\begin{equation}
\label{wfintegrals}
\begin{split}
h_{kl}&={\textstyle\int}\chi_k({\bf x})\{({-\hbar^2}/{2m})\nabla^2+V_{\textrm{ext}}({\bf x})\}\chi_l({\bf x})d^3x\\
V_{klmn}&={\textstyle\int}\chi_k({\bf x})\chi_l({\bf x}')V({\bf x},{\bf x}')\chi_m({\bf x})\chi_n({\bf x}')d^3xd^3x'
\end{split}
\end{equation}
with $k,l,m,n=1,2$ are matrix elements of the single-particle Hamiltonian $h({\bf x})$ and two-body interaction potential $V({\bf x},{\bf x}').$ Throughout this paper, the wavefunctions $\chi_1({\bf x})$ and $\chi_2({\bf x})$ are real-valued functions so that 
\begin{equation}
\begin{split}
V_{klmn}&=V_{lknm}=V_{mnkl}=V_{nmlk}=V_{mlkn}\\
&=V_{lmnk}=V_{knml}=V_{nklm}.
\end{split}
\end{equation}

This Hamiltonian is more general than that, for example, of Spekkens and Sipe \cite{Spekkens1999a} in that the external potential $V_{\textrm{ext}}$ is not assumed to be symmetric {\it{a priori}} since we anticipate the possibility for deformation of a single well into either a symmetric or asymmetric double well. Exclusion of certain nonlinear terms and the assumption of $N$-independent $h_{kl}$ and $V_{klmn},$ reduces (\ref{hamtwomode}) to a two-state Bose-Hubbard Hamiltonian. In the thermodynamic limit, Bose-Hubbard theory provides, {\it inter alia}, the standard model for the description of zero temperature quantum phase transitions \cite{Fisher1989a,Sachdev1999a} in atomic gases confined to optical lattices \cite{Jaksch1998b}.

With the Hamiltonian (\ref{hamtwomode}) and associated eigenstates (\ref{twomodestatesnl}), the many-body Schr\"odinger equation
\begin{equation}
\hat{H}|\Psi^N\rangle_\nu=E^\nu|\Psi^N\rangle_\nu,
\end{equation}
which is a linear equation, may be represented in the restricted Fock state basis (\ref{set}). This results in the set of $(N+1)\times(N+1)$ matrix equations
\begin{equation}
\label{Hammatrix}
\left[
\begin{array}{cccc}
H_{00}&H_{01}&\cdots&H_{0N}\\
H_{10}&H_{11}&\cdots&H_{1N}\\
\vdots&\vdots&\ddots&\vdots\\
H_{N0}&H_{N1}&\cdots&H_{NN}\\
\end{array}
\right]
\left[
\begin{array}{c}
C_0^\nu\\
C_1^\nu\\
\vdots\\
C_N^\nu\\
\end{array}
\right]
=
\left[
\begin{array}{c}
C_0^\nu\\
C_1^\nu\\
\vdots\\
C_N^\nu\\
\end{array}
\right]
E^\nu
\end{equation}
with matrix elements and overlap 
\begin{equation}
\begin{split}
H_{N'_1N_1}&=\langle N'_1,N-N'_1|\hat{H}|N_1,N-N_1\rangle\\
\delta_{N'_1N_1}&=\langle N'_1,N-N'_1|N_1,N-N_1\rangle
\end{split}
\end{equation}
for $N'_1,N_1=0,\ldots,N.$ Once the wavefunctions $\chi_1$ and $\chi_2$ are specified, the matrix elements $h_{kl}$ and $V_{klmn}$ are defined and can be used to build $H_{N'_1N_1}.$  Then, the eigenvalue equations (\ref{Hammatrix}) can, in principle, be solved to give the expansion coefficients $C_{N_1}^\nu$ and energies $E^\nu$ of the ground state and first $N$ excited states of the system. This computation may be repeated as the external potential $V_{\textrm{ext}}$ is deformed from a single well to a double well where the geometry encourages break up into two condensate fragments. Following Penrose and Onsager \cite{Penrose56}, a fragmented condensate may be characterized by the presence of at least two large eigenvalues of the single-particle reduced density matrix \cite{Lowdin55a,Leggett2001a}
\begin{equation}
\gamma^\nu({\bf x},{\bf x}')={_\nu\langle}\Psi^N|\hat{\Psi}^\dagger({\bf x})\hat{\Psi}({\bf x}')|\Psi^N\rangle_\nu.
\end{equation}


Within this model space, the full diagonalization of (\ref{Hammatrix}) has been found, in practice, to be achievable for systems with a large number of atoms. This is facilitated by the significant reduction of the Hamiltonian matrix in (\ref{Hammatrix}) to a pentadiagonal form \cite{Spekkens1999a}. Standard linear algebra routines \cite{lapack} are well suited for the resulting banded eigenvalue problem.


\subsection{Model Calculations}
\begin{figure}
\psfrag{E}[][]{{\large parameterized model energy ($\hbar\omega$)}}
\psfrag{alpha}[][]{{\large $\alpha$}}
\rotatebox{0}{\resizebox{!}{6cm}{\includegraphics{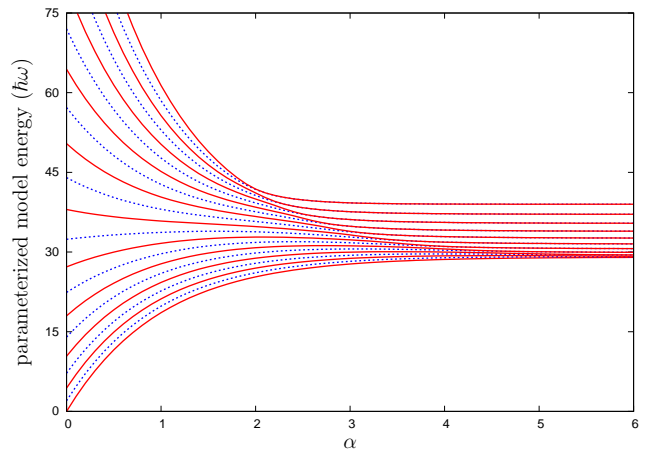}}}
\caption{\label{heidiplot}(Color online) Parameterized model ground and excited state energies for $g=0.1$ $\hbar\omega\cdot\beta^{3}$ and $N=20$ as a function of $\alpha.$ Note the energy level mergings and resulting ridge structure separating BEC (small $\alpha$) and fragmented states (large $\alpha$). The Fock states below the ridge are delocalized and nondegenerate while the states above the ridge are localized and doubly degenerate. The alternation of line style has been chosen to aid in visualization.}
\end{figure}

The energies of the ground and first $N$ excited states of the Hamiltonian (\ref{hamtwomode}) have been mapped out as a function of the single dimensionless parameter $\alpha$ in correlation diagrams by Reinhardt and Perry \cite{Reinhardt2003a} as illustrated in Figure \ref{heidiplot}. In \cite{Reinhardt2003a}, it has been assumed that the single-particle energies and matrix elements of the two-body potential $V,$ in the standard contact interaction approximation $V({\bf x},{\bf x}')=g\delta({\bf x}-{\bf x}'),$ can be parameterized according to \footnote{The specific parameterization of the matrix elements of $h$ and $V$ given in \cite{Reinhardt2003a,Mahmud05} has been misprinted. Rather, the correct parameterization, which is associated with the energy level correlation diagram in Figure \ref{heidiplot}, is given in (\ref{alpharules}). We explicitly indicate the energy units $\hbar\omega$ in these formulae whereas \cite{Reinhardt2003a,Mahmud2003a,Mahmud05} do not.}
\begin{equation}
\label{alpharules}
\begin{array}{c}
V_{kkkk}=h_{kk}=\hbar\omega\\
V_{kkkl}=h_{kl}=-\hbar\omega\exp(-\alpha)\\
V_{kkll}=\hbar\omega\exp(-\alpha)
\end{array}
\end{equation}
in terms of the harmonic oscillator energy $\hbar\omega$ and length $\beta=\sqrt{\hbar/m\omega}$ for a symmetric double well trapping potential with $k\neq l=1,2.$ Variation of the parameter $\alpha$ allows for a continuous change between strong tunneling (small $\alpha$) and weak tunneling (large $\alpha$) regimes. Within this simple ansatz, where none of the parameters depend upon the particle number $N,$ the wavefunctions $\chi_1$ and $\chi_2$ are not specified. The ridge structure in Figure \ref{heidiplot} shows the boundary between simple BEC for small $\alpha$ and fragmented BEC for large $\alpha.$ When working in the Fock basis (\ref{2fs}) where $N_1$ and $N_2$ are the number of atoms localized in the left and right wells of a double well potential, it has been found in \cite{Mahmud2003a} that the distribution of Fock states contributing to the ground state below the ridge is binomial in form, while the distribution of Fock states above the ridge reveals the existence of macroscopic quantum superposition states. The study of the correlation diagrams in Figure \ref{heidiplot} has led to the prediction that many interesting highly excited states, such as Schr\"odinger cats, exist in the weak tunneling regime and may be created by phase engineering \cite{Mahmud2003a,Mahmud05}. 
\begin{figure}
\psfrag{E}[][]{{\large Schr\"odinger model energy ($\hbar\omega$)}}
\psfrag{alpha}[][]{{\large barrier height ($\hbar\omega$)}}
\rotatebox{0}{\resizebox{!}{6cm}{\includegraphics{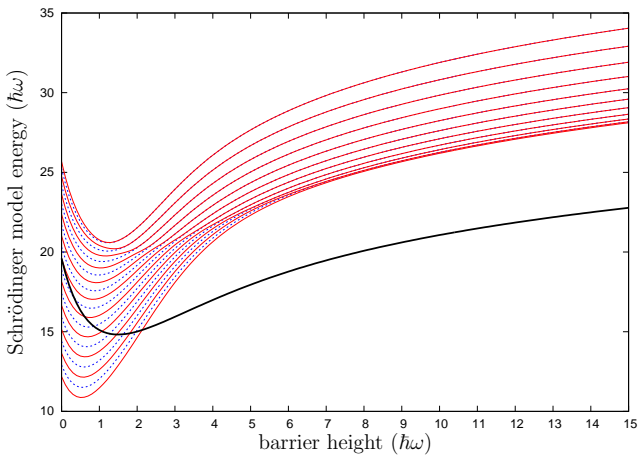}}}
\caption{\label{sch}(Color online) Schr\"odinger model ground and excited state energies for $g=0.1$ $\hbar\omega\cdot\beta^{3}$ and $N=20$ as a function of barrier height. Note the qualitative similarities of this energy level correlation diagram to that of Figure \ref{heidiplot}. The solid curve beginning at 20 $\hbar\omega$ is the single-particle Schr\"odinger energy while the remaining curves are the eigenvalues of the Hamiltonian (\ref{hamtwomode}) built from the Schr\"odinger solutions at each barrier height. Note that the Schr\"odinger energy $N_1\varepsilon_1+N_2\varepsilon_2$ differs from the many-body eigenvalues $E^\nu$ because it includes no atom-atom interaction. The alternation of line style has been chosen to aid in visualization.}
\end{figure}

Other authors \cite{Spekkens1999a} opt to approximate the single-particle wavefunctions $\chi_1$ and $\chi_2$ for a given barrier height by solutions of the single-particle Schr\"odinger equation 
\begin{equation}
\label{1se}
h({\bf x})\chi_k({\bf x})=\varepsilon_k\chi_k({\bf x})
\end{equation}
where $\varepsilon_k\equiv h_{kk}/\langle\chi_k|\chi_k\rangle$ is the single-particle Schr\"odinger energy for state $k$ and where the wavefunctions are independent of the particle number $N$ and value of $g.$ The two-state Hamiltonian (\ref{hamtwomode}), which does depend upon $N$ and upon $g$ in the contact interaction approximation, is then diagonalized. Figure \ref{sch} presents the associated energy eigenvalues as a function of barrier height, where $\chi_1$ and $\chi_2$ are computed from (\ref{1se}). A qualitatively similar energy correlation diagram exists with ridge structure marking the phase transition between nondegenerate states and doubly degenerate states. This approach is justifiable for weakly interacting atomic gases where the effect of atomic interactions on the shape of the wavefunctions is small, however, it breaks down wherever atom-atom interactions are important enough to affect the value of the parameters or matrix elements themselves. This is the case in the experiments discussed in \cite{Shin2004a,Saba05,Anker04,Albiez04}, where Thomas-Fermi mean field effects dominate the shape of the atomic wavefunctions in a strongly $N$-dependent manner.

\subsection{Lack of Mean Field Effects in the Single-Particle Wavefunctions}
Diagonalization of the Hamiltonian (\ref{hamtwomode}) within the basis of restricted Fock states (\ref{set}), furnishes a basis representation of the ground state and $N$ excited states due to exchanges of atoms between the two single-particle states $|\chi_1\rangle$ and $|\chi_2\rangle.$ As no equations have been derived for these states, their functional form is not specified by this approach and approximate models have been chosen that either define the $\chi_1$ and $\chi_2$ as solutions of the single-particle Schr\"odinger equation or parameterize the Hamiltonian matrix elements $h_{kl}$ and $V_{klmn}$ directly. Neither method takes into account the effects of the condensate mean field on the shape of the single-particle wavefunctions, which becomes more important as the interaction strength between the constituent atoms becomes larger. We will demonstrate, in Section IV, how to build atomic interactions into the underlying single-particle states that enter the matrix elements in the many-body Hamiltonian (\ref{hamtwomode}). But first, it is important to discuss mean field theory by itself for identical bosons.

\section{Review of Mean Field Theory for Identical Bosons}
Single-particle wavefunctions $\chi_k$ were first introduced in the Fock space approach of the previous section. Within that model, no equations were developed to determine these functions. In this section, we derive a set of equations by minimizing the energy associated with (\ref{hamtwomode}) to dictate the functional form of the two $\chi_k.$ The equations that arise are the bosonic Hartree-Fock equations.

\subsection{Restriction to One Single-Particle State}
The mean field theory in which all $N$ atoms occupy the same single-particle state $|\chi\rangle=\hat{b}^\dagger|\textrm{vac}\rangle$ is known as the Gross-Pitaevskii (GP) theory. The many-body wavefunction $\Psi^N$ restricted to the Fock state $|N\rangle=(\hat b^\dagger)^{N}|{\textrm{vac}}\rangle/\sqrt{N!}$ is approximated by the product 
\begin{equation}
\Psi^H(1,\ldots,N)=\chi(1)\chi(2)\cdots\chi(N)
\end{equation}
of single-particle wavefunctions $\chi.$ This particular type of product is also called a Hartree product since it involves no symmetrization whatsoever. Variation of the expectation value of the many-body Hamiltonian (\ref{hamfockpseudopotential}) with $\hat{\Psi}=\chi\hat{b}$ in $|\Psi^H\rangle\equiv|N\rangle$ with respect to $\chi$ and subject to the constraint that $\chi$ is normalized to unity leads to the GP equation
\begin{equation}
\label{gpe1}
\{(-{\hbar^2}/{2m})\nabla^2+V_{\textrm{ext}}+g(N-1)|\chi|^2\}\chi=\mu\chi
\end{equation}
provided that the contact interaction approximation has been made. The chemical potential $\mu$ enters (\ref{gpe1}) as a Lagrange multiplier which enforces the normalization $\langle\chi|\chi\rangle=1.$ While this approximation provides an appropriate mean field description of the simple BEC, it is not flexible enough to characterize the break up of a single condensate into multiple fragments, where potentially several single-particle wavefunctions are macroscopically occupied.

\subsection{Restriction to Two Single-Particle States}
A mean field theory, which generalizes the GP (or Hartree) ansatz by adding a second single-particle state, is the bosonic Hartree-Fock (BHF) theory. The BHF ansatz rests on approximating the many-body wavefunction for $N$ bosons as a {\it symmetric} product of the two single-particle wavefunctions $\chi_1$ and $\chi_2$ \cite{Esry1997c,Cederb04}. That is 
\begin{equation}
\begin{split}
\label{hfa}
\Psi^{\textrm{BHF}}&(1,\ldots,N)\\
&={\cal S}\{\chi_1(1)\cdots\chi_1(N'_1)\chi_2(N'_1+1)\cdots\chi_2(N'_1+N'_2)\},
\end{split}
\end{equation}
where ${\cal S}=(\sqrt{N!})^{-1}\sum_PP$ is the symmetrization operator and $P$ is an operator that permutes the atomic coordinates. We place primes on the occupation numbers $N'_1$ and $N'_2$ for reasons that will become evident in Section IV. This BHF wavefunction is also called a single permanental wavefunction in contrast to the single determinantal wavefunction for fermions built from an {\it antisymmetric} product of single-particle wavefunctions. If we had omitted the symmetrization operator $\cal S$ all together in (\ref{hfa}), then we would have a two-single-particle state Hartree or GP theory \cite{Faddeev}. Two-state Hartree theory provides an alternative mean field theory that neglects the quantum-mechanical exchange interaction associated with identical particles. We note that the ansatz (\ref{hfa}) has been explored in a different context in \cite{Alon05}.

The coupled BHF equations may be determined by variation of the expectation value of the functional
\begin{equation}
\label{funcK}
\hat{K}[\chi_1,\chi_2]=\hat{H}-\sum_{kl=1,2}\mu_{kl}\hat{N}'_k(\Delta_{kl}-\delta_{kl})
\end{equation}
in the BHF state
\begin{equation}
\label{2hfs}
|\Psi^{\textrm{BHF}}\rangle\equiv|N'_1,N'_2\rangle=(\hat b_1^\dagger)^{N'_1}(\hat b_2^\dagger)^{N'_2}|{\textrm{vac}}\rangle/\sqrt{N'_1!N'_2!}
\end{equation}
with respect to the two wavefunctions $\chi_1$ and $\chi_2,$ where $\hat H$ is the many-body Hamiltonian (\ref{hamtwomode}) restricted to the model space of two single-particle states, $\hat N={\hat N}'_1+{\hat N}'_2$ is the total particle number operator, and $\Delta_{kl}\equiv\langle\chi_k|\chi_l\rangle$ are the matrix elements of the wavefunction overlap $\Delta.$ The second term on the right hand side of this equation adds Lagrange multipliers $\mu_{kl}$ whose purpose is to constrain the single-particle wavefunctions to be orthonormal. This leads to the coupled two-single-particle state BHF equations
\begin{equation}
\begin{split}
\label{coupledgpe}
h\chi_1+(N'_1-1)\Gamma_{1}\chi_1+N'_2[{\cal J}_{2}+{\cal K}_{2}]\chi_1&=\mu_{11}\chi_1+\mu_{12}\chi_2\\
h\chi_2+(N'_2-1)\Gamma_{2}\chi_2+N'_1[{\cal J}_{1}+{\cal K}_{1}]\chi_2&=\mu_{21}\chi_1+\mu_{22}\chi_2,
\end{split}
\end{equation}
where $h({\bf x})=({-\hbar^2}/{2m})\nabla^2+V_{\textrm{ext}}({\bf x})$ is the sum of kinetic energy and external trapping potential, $\Gamma_{k}$ accounts for the interaction of one atom in the $k$th single-particle state with the mean field of $N'_k-1$ other atoms in the same state, ${\cal J}_{l}$ is the direct interaction between a single atom in the $k$th single-particle state and the mean field of $N'_{l}$ atoms in the $l$th ($l\neq k$) single-particle state, and ${\cal K}_{l}$ is the exchange interaction between states $k$ and $l$ which arises due to the symmetrization of the BHF wavefunction. These equations have already been derived by others \cite{Esry1997c,Cederbaum2003a,Cederb04} and have been further extended to treat identical bosons in arbitrarily many single-particle states \cite{Alon05}. Similar equations have been for derived for {\it distinguishable} multicomponent (spinor) BECs \cite{Ho1996a,Esry1997a,Ballagh1997a,Graham1998b}, however, it is important to note that they are not the same as the BHF equations (\ref{coupledgpe}), which describe {\it identical} bosons.

The diagonal Lagrange multipliers $\mu_{kk}$ in (\ref{funcK}) ensure the proper normalization of the single-particle wavefunctions while the off-diagonal $\mu_{kl}$ enforce their orthogonality. If the external potential $V_{\textrm{ext}}$ is symmetric, then the off-diagonal Lagrange multipliers are not necessary as the wavefunctions are automatically spatially orthogonal by symmetry. The $\mu_{kl}$ cannot in general be removed by unitary transformation as in the fermionic case \cite{Cederbaum2003a,Cederb04}. Consequently, arbitrary linear combinations of $\chi_1$ and $\chi_2$ are not solutions of the BHF equations (\ref{coupledgpe}). Koopmans' theorem \cite{Koopmans34,Ostlund,Levine} is satisfied for the diagonal multipliers. Therefore $\mu_{11}=E^{\textrm{BHF}}[N'_1,N'_2]-E^{\textrm{BHF}}[N'_1-1,N'_2]$ and $\mu_{22}=E^{\textrm{BHF}}[N'_1,N'_2]-E^{\textrm{BHF}}[N'_1,N'_2-1]$ inherit the roles of chemical potentials \cite{Cederbaum2003a,Cederb04}, where the BHF energy is given by
\begin{equation}
\begin{split}
\label{BHFE}
E^{\textrm{BHF}}&=\sum_{k=1,2}N'_kh_{kk}+(1/2)\sum_{k=1,2}N'_k(N'_k-1)V_{kkkk}\\
&\ \ \ +(1/2)\sum_{k\neq l=1,2}N'_kN'_l[V_{klkl}+V_{kllk}], 
\end{split}
\end{equation}
and where $h_{kl}$ and $V_{klmn}$ are matrix elements of the single-particle Hamiltonian $h({\bf x})$ and two-body interaction potential $V({\bf x},{\bf x}')$ in (\ref{wfintegrals}). The direct and exchange integrals in (\ref{coupledgpe}) are defined as
\begin{equation}
\begin{split}
\Gamma_{k}({\bf x})\chi_k({\bf x})&={\textstyle\int_V}[\chi_k({\bf x}')V({\bf x},{\bf x}')\chi_k({\bf x}')]\chi_k({\bf x})d^3x'\\
{\cal J}_{l}({\bf x})\chi_{k}({\bf x})&={\textstyle\int_V}[\chi_{l}({\bf x}')V({\bf x},{\bf x}')\chi_{l}({\bf x}')]\chi_{k}({\bf x})d^3x'\\
{\cal K}_{l}({\bf x})\chi_{k}({\bf x})&={\textstyle\int_V}[\chi_{l}({\bf x}')V({\bf x},{\bf x}')\chi_{k}({\bf x}')]\chi_{l}({\bf x})d^3x'
\end{split}
\end{equation}
for $k\neq l$ and $k,l=1,2.$ Note that the potential $\Gamma_{k}$ is a direct interaction that arises only for bosons. There is no analogous term for fermions due to Pauli exclusion.

\begin{figure}
\psfrag{CI and MF energy}[][]{{\large $E^{\textrm{BHF}}$ ($\hbar\omega$)}}
\psfrag{N1/N}[][]{{\large $N'_1/N$}}
\psfrag{GP1es}[][]{GP 1es\hspace{.3cm}}
\psfrag{BHF}[][]{BHF}
\psfrag{GPgs}[][]{GP gs\hspace{.3cm}}
\rotatebox{0}{\resizebox{!}{6cm}{\includegraphics{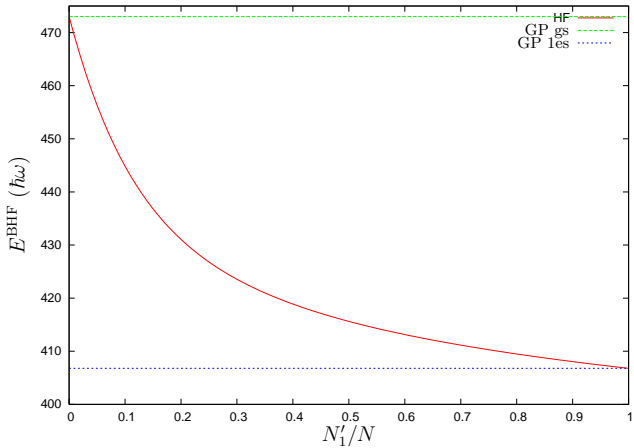}}}
\caption{\label{2m-mfe}(Color online) The BHF energy reduces to that of the GP ground state when $N'_1/N=1$ and to that of the first GP excited state when $N'_1/N=0.$ This corresponds to all particles in the symmetric GP ground state and the first antisymmetric GP excited state respectively. A dimensionless interaction strength of $\alpha_{\textrm{Q1D}}=40$ [see (\ref{gq1d})] for 100 atoms in a single well potential was used in this calculation of $E^{\textrm{BHF}}.$}
\end{figure}

The BHF theory reduces to the GP theory in the extreme limit of only one occupied single-particle state, {\it i.e.}, when $N'_1=N$ and $N'_2=0$ or vice versa as demonstrated in Figure \ref{2m-mfe}. 
However, unlike the GP equation (\ref{gpe1}), the BHF equations (\ref{coupledgpe}) offer additional freedom in that they accommodate the possibility of various numbers of atoms in each of two single-particle states. By choosing $N'_1$ atoms to be in the first state and $N'_2$ atoms to be in the second, a set of BHF equations corresponding to that particular arrangement is obtained. Each arrangement is associated with a symmetrized BHF single-permanental wavefunction like (\ref{hfa}). Where BHF equations specify the underlying single-particle wavefunctions $\chi_1$ and $\chi_2,$ we call the Fock state $|\Psi^{\textrm{BHF}}\rangle\equiv|N'_1,N'_2\rangle$ a {\it single configuration.} These Fock states are distinguished from those in Section II, which were written without primes as $|N_1,N_2\rangle,$ because the single-particle states that make up each single configuration are determined by solving BHF equations. It is then possible to vary the number of particles in each of the single-particle states to find the lowest energy single configuration for a particular trap geometry. This flexibility gives rise to two very different physical meanings for $\chi_1$ and $\chi_2.$ In the first, $N'_1\sim N$ and there are a relatively small number of atoms in $|\chi_2\rangle.$ In this regime, $|\chi_1\rangle$ is a condensate state and $|\chi_2\rangle$ is a single-particle excited state. The energetic cost of making excitations from $|\chi_1\rangle$ to $|\chi_2\rangle$ is macroscopically large due to the effect of bosonic amplification \cite{Huang1957a}. In the second, fragmented, regime, $N'_1$ and $N'_2$ are {\it both} on the order of $N.$ In this case, both single-particle states are condensate states. Single-atom excitations between $|\chi_1\rangle$ and $|\chi_2\rangle$ also cost a macroscopic amount of energy, but now there is also a macroscopic gain of energy as the atom is added to a second macroscopically occupied state \cite{Reinhardt2003a}. This distinction, which is quite important, will be elaborated on in Section V.


\subsection{Numerical Integration Method}
We numerically solve the coupled BHF equations using a fast Fourier transform based pseudospectral grid method \cite{fftw,Gottlieb} in quasi-one dimension \cite{Carr00,Olshanii1998a}. Note that quasi-one dimension does not mean one dimension but rather that the variation of the BEC density is negligible in the two transverse dimensions and a separation of variables is permissible. Using this approach, the equations are expanded onto a discrete Fourier sine basis with $2^8$ fixed grid points \footnote{We have also performed calculations with $2^9$ and $2^{10}$ fixed grid points but our results have adequately converged with only $2^8$ points.}, which satisfies the proper boundary conditions, and the expansion coefficients are variationally optimized. Rather than solving the time-independent equations (\ref{coupledgpe}) self-consistently, we work with their time-dependent version, where $\mu_{11}$ and $\mu_{22}$ are replaced by $i\hbar(d/dt)$ \cite{Esry1997a}. We then solve the time-dependent BHF equations by the method of steepest descents in imaginary time \cite{Dalfovo96}. That is, we employ the Wick time rotation $t\to\tau=it,$ which takes $i\hbar(d/dt)$ to $-\hbar(d/d\tau),$ and integrate the pair of coupled nonlinear diffusion equations
\begin{equation}
\begin{split}
\label{t-2modeH}
-\hbar(d/d\tau)\chi_1&={\cal F}_1\chi_1-\mu_{12}\chi_2\\
-\hbar(d/d\tau)\chi_2&={\cal F}_2\chi_2-\mu_{21}\chi_1
\end{split}
\end{equation}
as an initial value problem in $\tau,$ where ${\cal F}_k$ is the boson Fock operator for the $k$th single-particle state ($k=1,2$),
\begin{equation}
\label{fop}
{\cal F}_k({\bf x})=h({\bf x})+(N'_k-1)\Gamma_{k}({\bf x})+N'_{l}[{\cal J}_{l}({\bf x})+{\cal K}_{l}({\bf x})].
\end{equation}
The resulting equations are well-defined once the two-body interaction potential $V$ is identified. As was done in Section II, we make use of the contact potential $V({\bf x},{\bf x}')=(4\pi\hbar^2a/m)\delta({\bf x}-{\bf x}').$ Following the argument of Esry {\it et. al.} \cite{Esry1997a}, the renormalized $S$-wave scattering length $a$ is taken from multichannel $T$-matrix calculations using {\it symmetrized} two-body wavefunctions. Thus, the contact interaction approximation effects the replacement of the symmetric combination of two-body matrix elements in the BHF energy (\ref{BHFE}) with a single contact potential and not each matrix element separately. In symbols that is
\begin{equation}
V_{klkl}+V_{kllk}\equiv\langle kl|V|kl+lk\rangle\to(4\pi\hbar^2a/m)\delta({\bf x}-{\bf x}').
\end{equation}
This identification differs by a factor of two from \cite{Cederbaum2003a,Cederb04,Alon05}; see \footnote{In a one-dimensional model of interacting bosons \cite{Lieb63}, the two-body potential $V(x,x')=g\delta(x-x')$ is a real interaction potential and is {\it not} an effective potential. With this choice for $V,$ the interaction terms $N'_l[{\cal J}_l+{\cal K}_l]\chi_k$ in the BHF equations (\ref{coupledgpe}) would become $2gN'_l|\chi_l|^2\chi_k$ \cite{Cederbaum2003a,Cederb04,Alon05} rather than $gN'_l|\chi_l|^2\chi_k$ as we have in (\ref{pcoupledgpe}). This is due to the fact that $\langle kl|V|kl+lk\rangle$ is no longer replaced by an effective potential, but rather each matrix element $\langle kl|V|kl\rangle$ and $\langle kl|V|lk\rangle$ contributes separately to the equations since $V$ is a real scattering potential.}, \footnote{It has been found that BHF equations like those presented in \cite{Cederbaum2003a,Cederb04,Alon05}, but which differ from (\ref{pcoupledgpe}) by a factor of two in the cross interaction terms, yield MCBHF ground state and low lying excited state energies that are higher than those obtained with a factor of one. As the associated underlying BHF wavefunctions provide only a basis for the matrix elements in the many-body Hamiltonian (\ref{hamtwomode}), we choose to perform all MCBHF calculations with (\ref{pcoupledgpe}) as opposed to the BHF equations appearing in \cite{Cederbaum2003a,Cederb04,Alon05}.} and the discussion following (\ref{Eset2}). The resulting BHF equations (\ref{coupledgpe}) are \cite{Esry1997a,Esry1997c}
\begin{equation}
\begin{split}
\label{pcoupledgpe}
\{h+g(N'_1-1)|\chi_1|^2+gN'_2|\chi_2|^2\}\chi_1&=\mu_{11}\chi_1+\mu_{12}\chi_2\\
\{h+g(N'_2-1)|\chi_2|^2+gN'_1|\chi_1|^2\}\chi_2&=\mu_{21}\chi_1+\mu_{22}\chi_2.
\end{split}
\end{equation}


We choose as an initial condition for the single-particle wavefunction $\chi_1$ the square root of the Thomas-Fermi density $\rho^{\textrm{TF}}({\bf x})=[\mu-V_{\textrm{ext}}({\bf x})]/g(N-1)$ at a particular barrier height, chemical potential, and value of coupling constant. For a symmetric trapping potential, the second wavefunction $\chi_2$ is taken to be antisymmetric to $\chi_1$ and the off-diagonal Lagrange multipliers are unnecessary. Otherwise, $\chi_2$ need only be orthogonal to $\chi_1$ but constraints are needed to maintain orthogonality. Both wavefunctions are initially normalized so that $\|\chi_1(0)\|=\|\chi_2(0)\|=1$ and are real-valued. Thus, the overlap matrix elements $\Delta_{kl}=\delta_{kl}$ initially.  We then employ the standard relaxation approach, {\it i.e.}, we subtract a guess $\bar\mu$ of the ground state chemical potential from the Fock operator and allow the system to time evolve. The wavefunctions relax according to
\begin{equation}
\label{nldes}
\chi_k(\tau)\approx\sum_\mu\exp(-[\mu-\bar{\mu}_{kk}]\tau/\hbar)\chi_k^\mu(0)c_\mu
\end{equation}
for $k=1,2,$ where the expansion coefficients $c_\mu$ are projections of the evolving state $|\chi_k(\tau)\rangle$ onto the stationary basis $|\chi_k^\mu(\tau)\rangle.$ Eventually, all excited states decay away after repetition of this procedure together with intermittent renormalization. If the wavefunctions are initially orthogonal and share the symmetry of the trapping potential, then the wavefunctions that remain are the symmetric and antisymmetric solutions which minimize $E^{\textrm{BHF}}$ for each configuration.

Whenever the external potential $V_{\textrm{ext}}$ is asymmetric, off-diagonal Lagrange multipliers must be introduced to keep the single-particle wavefunctions orthogonal to each other throughout the evolution. The proper choice for the $\mu_{kl}$ $(k\neq l)$ will ensure that $(d/d\tau)\Delta_{kl}(\mu_{kl})=0$ for all time $\tau.$ The appropriate $\mu_{kl}$ $(k\neq l)$ are found by multiplying the first equation in (\ref{t-2modeH}) by $\chi_2$ and the second equation in (\ref{t-2modeH}) by $\chi_1,$ adding the two equations together, and then performing a volume integral. One then arrives at the following expression for the time derivative of the off-diagonal overlap
\begin{equation}
\begin{split}
\label{delta12}
-\hbar(d/d\tau)\Delta_{12}&=\langle\chi_2|{\cal F}_1-\bar\mu_{11}|\chi_1\rangle-\mu_{12}\langle\chi_2|\chi_2\rangle\\
&\ \ \ +\langle\chi_1|{\cal F}_2-\bar\mu_{22}|\chi_2\rangle-\mu_{21}\langle\chi_1|\chi_1\rangle,
\end{split}
\end{equation}
where the $\bar{\mu}_{kk}$ are intermediate guesses of the ground state chemical potentials associated with $\chi_1$ and $\chi_2.$ The time derivative of $\Delta_{21}$ is the same, as $\Delta^T=\Delta.$ With (\ref{delta12}) and the relation $N'_1\mu_{12}=N'_2\mu_{21}$ \cite{Cederb04}, values for $\mu_{12}$ and $\mu_{21}$ can be chosen so that the right hand side of (\ref{delta12}) equals zero. Those values are
\begin{equation}
\begin{split}
\label{mu12}
\mu_{12}&=\frac{\langle\chi_2|{\cal F}_1-\bar\mu_{11}|\chi_1\rangle+\langle\chi_1|{\cal F}_2-\bar\mu_{22}|\chi_2\rangle}{(N'_1/N'_2)\langle\chi_1|\chi_1\rangle+\langle\chi_2|\chi_2\rangle}\\
\mu_{21}&=\frac{\langle\chi_2|{\cal F}_1-\bar\mu_{11}|\chi_1\rangle+\langle\chi_1|{\cal F}_2-\bar\mu_{22}|\chi_2\rangle}{\langle\chi_1|\chi_1\rangle+(N'_2/N'_1)\langle\chi_2|\chi_2\rangle}.
\end{split}
\end{equation}
With this choice, the time derivatives of the off-diagonal matrix elements of $\Delta$ are zero. Therefore, if the single-particle wavefunctions are initially orthogonal, then they will stay orthogonal for all time $\tau$ regardless of trap symmetry. For each choice of BHF configuration $|N'_1,N'_2\rangle,$ the relaxation algorithm will then find the associated two orthogonal wavefunctions which minimize the energy (\ref{BHFE}). This is the essence of our integration scheme.

\subsection{Hartree-Fock and Schr\"odinger Wavefunctions}
\begin{figure*}
\psfrag{psi}[][]{{\large $\chi_1$ and $\chi_2$ $(\beta^{-1/2})$}}
\psfrag{psi1}[][]{\rotatebox{180}{{\large $V_{\textrm{ext}}$ $(\hbar\omega)$}}}
\psfrag{x}[][]{{\large $z$ $(\beta)$}}
\psfrag{mode 1}[][]{$\chi_1$\hspace{-0.3cm}}
\psfrag{mode 2}[][]{$\chi_2$\hspace{-0.3cm}}
\psfrag{TF}[][]{TF\hspace{0.3cm}}
\psfrag{Vex}[][]{$V_{\textrm{ext}}\hspace{0.3cm}$}
\rotatebox{0}{\resizebox{!}{12.9cm}{\includegraphics{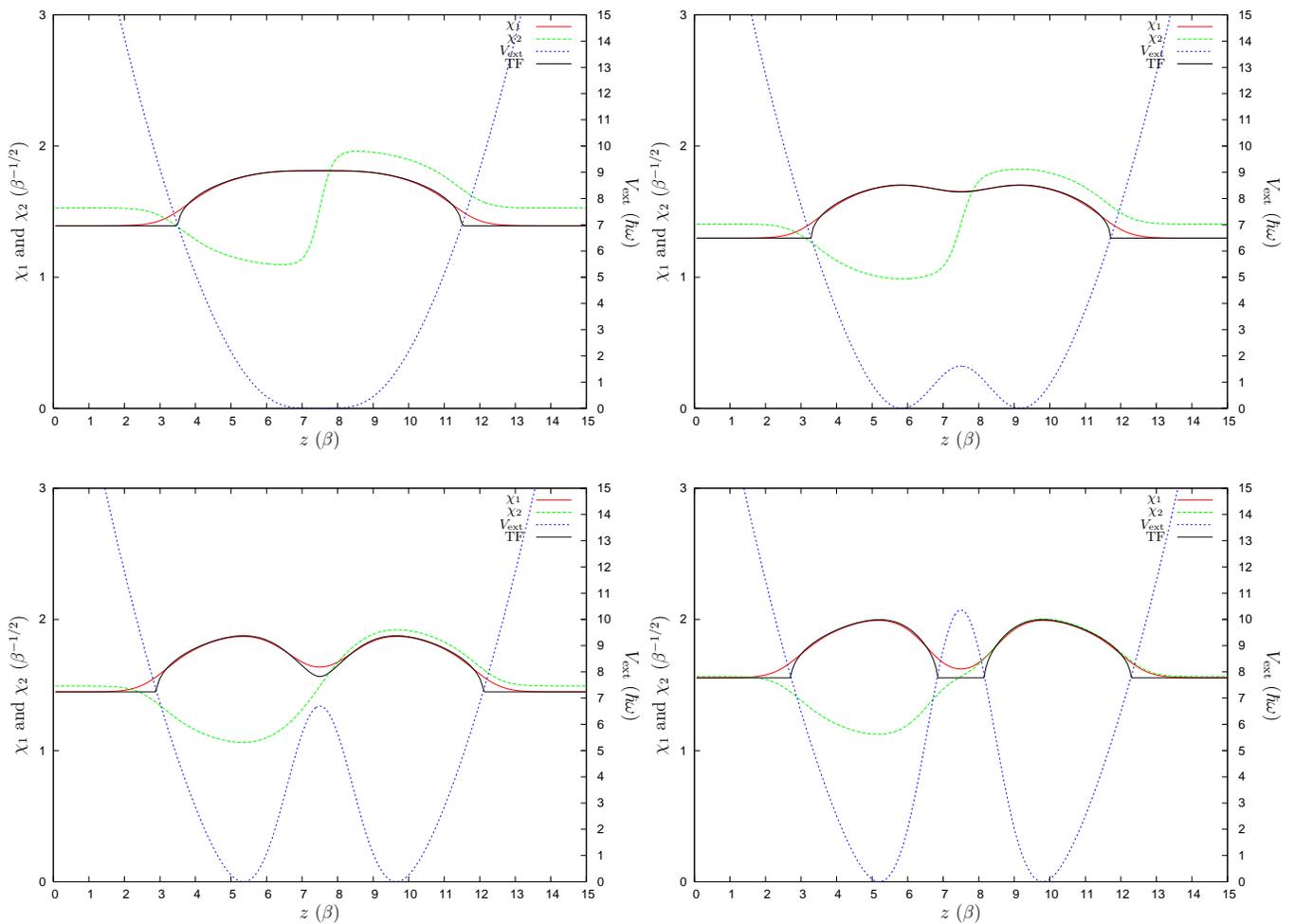}}}
\caption{\label{hfwffig}(Color online) BHF single-particle wavefunctions $\chi_1$ and $\chi_2$ versus coordinate $z$ at four different barrier heights 0, 3, 9, 13 $\hbar\omega.$ A dimensionless interaction strength of $\alpha_{\textrm{Q1D}}=40$ [see (\ref{gq1d})] was used. In each panel, $\chi_1$ corresponds to the BHF configuration $|N,0\rangle,$ while $\chi_2$ corresponds to the BHF configuration $|0,N\rangle.$ The BHF energies associated with $|N,0\rangle$ and $|0,N\rangle$ at zero barrier height appear in Figure \ref{2m-mfe}. Both wavefunctions have been set at the chemical potentials $\mu_{11}$ and $\mu_{22}$ associated with each configuration. At each barrier height, we plot the corresponding Thomas-Fermi wavefunction as a solid black curve.}
\end{figure*}

\begin{figure*}
\psfrag{psi}[][]{{\large $\chi_1$ and $\chi_2$ $(\beta^{-1/2})$}}
\psfrag{psi1}[][]{\rotatebox{180}{{\large $V_{\textrm{ext}}$ $(\hbar\omega)$}}}
\psfrag{x}[][]{{\large $z$ $(\beta)$}}
\psfrag{mode 1}[][]{$\chi_1$\hspace{-0.3cm}}
\psfrag{mode 2}[][]{$\chi_2$\hspace{-0.3cm}}
\psfrag{TF}[][]{TF\hspace{0.3cm}}
\psfrag{Vex}[][]{$V_{\textrm{ext}}\hspace{0.3cm}$}
\rotatebox{0}{\resizebox{!}{12.9cm}{\includegraphics{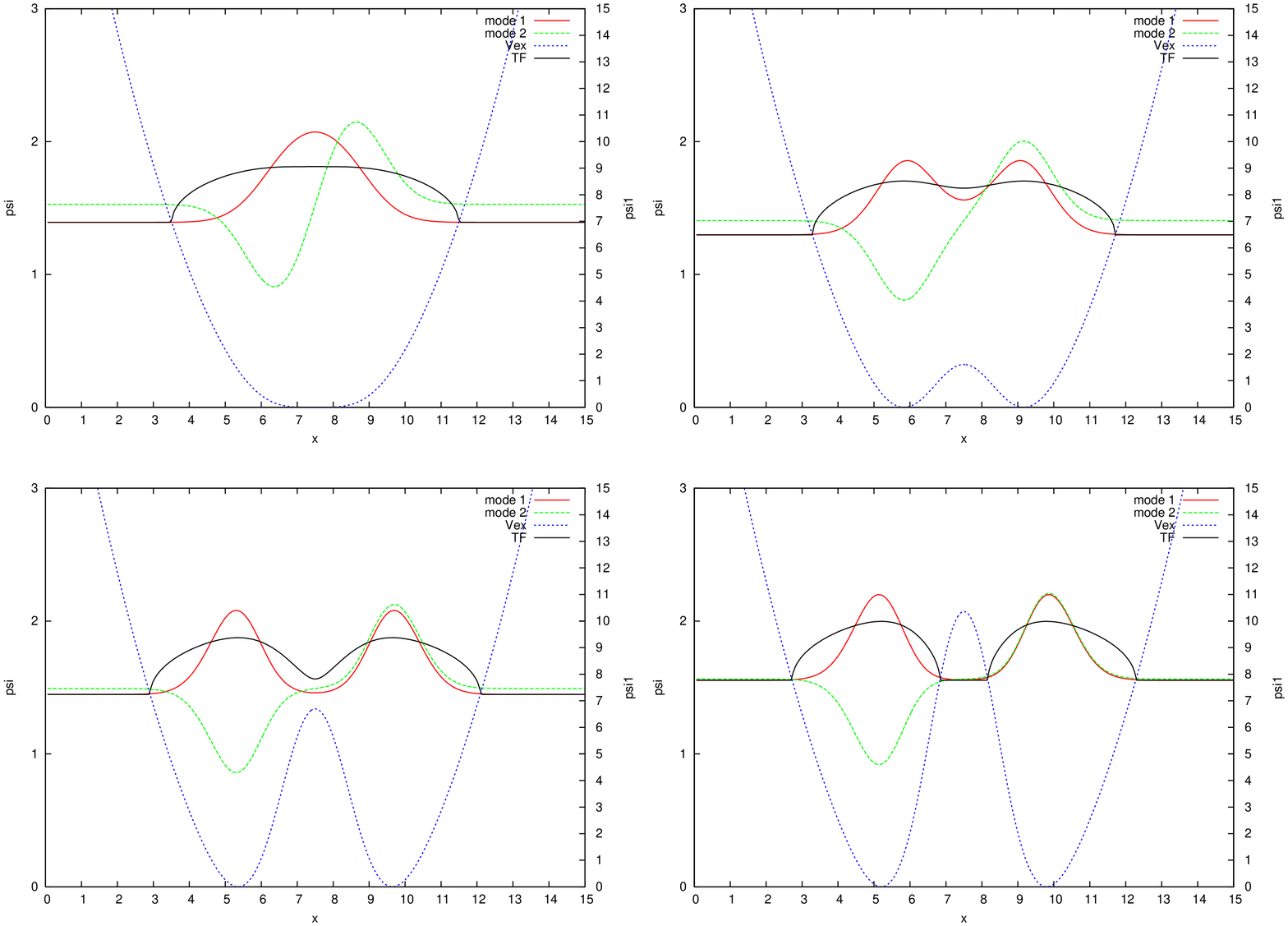}}}
\caption{\label{hfwffig_sch}(Color online) Schr\"odinger single-particle wavefunctions $\chi_1$ and $\chi_2$ versus coordinate $z$ at four different barrier heights 0, 3, 9, 13 $\hbar\omega.$ These solutions have no dependence on the number of atoms in each single-particle state. Both wavefunctions have been set at the same BHF chemical potentials used in Figure \ref{hfwffig}. At each barrier height, we plot the Thomas-Fermi wavefunctions of Figure \ref{hfwffig} as a solid black curve. Little resemblance is seen between the Schr\"odinger and Thomas-Fermi results.}
\end{figure*}

Figure \ref{hfwffig} displays the BHF single-particle wavefunctions $\chi_1$ and $\chi_2,$ which are obtained by solving (\ref{pcoupledgpe}) in quasi-one dimension at four different barrier heights 0, 3, 9, 13 $\hbar\omega.$ In each panel, $\chi_1$ is associated with the BHF configuration $|N,0\rangle$ and $\chi_2$ is associated with the BHF configuration $|0,N\rangle.$ In these extreme configurations, the single-particle BHF wavefunctions reduce to the GP ground and first excited wavefunctions.    
The configuration $|N,0\rangle$ minimizes $E^{\textrm{BHF}}$ at each barrier height. Figure \ref{hfwffig_sch} displays the solutions of the single-particle Schr\"odinger equation $h\chi_k=\varepsilon_k\chi_k$ at the same four barrier heights 0, 3, 9, 13 $\hbar\omega.$ Since the Schr\"odinger equation includes no atom-atom interaction, the Schr\"odinger solutions have no dependence on the number of particles in each single-particle state. For 100 atoms with a dimensionless interaction strength of $\alpha_{\textrm{Q1D}}=40$ [see (\ref{gq1d})], it is seen that the BHF wavefunctions follow the Thomas-Fermi result quite well whereas the Schr\"odinger wavefunctions bear little resemblance to either BHF or Thomas-Fermi solutions.

\subsection{Lack of Correlation Effects}
The solutions $\chi_1$ and $\chi_2$ of the BHF equations (\ref{coupledgpe}) include the effects of the condensate mean field on their shape. An appropriate symmetrized product of these wavefunctions results in a BHF wavefunction like that in (\ref{hfa}). Unlike in the single determinantal case where the wavefunction is defined only up to unitary transformations of its constituent single-particle wavefunctions \cite{Thouless60}, the BHF single permanental wavefunction seemingly corresponds uniquely to a single configuration. That is, the single permanent
\begin{equation}
{\cal S}\{\chi_1(1)\cdots\chi_1(N'_1)\chi_2(N'_1+1)\cdots\chi_2(N'_1+N'_2)\}
\end{equation}
is in one-to-one correspondence with the single configuration $|N'_1,N'_2\rangle,$ while another BHF permanent 
\begin{equation}
{\cal S}\{\chi_1(1)\cdots\chi_1(N''_1)\chi_2(N''_1+1)\cdots\chi_2(N''_1+N''_2)\}
\end{equation}
is in one-to-one correspondence with the single configuration $|N''_1,N''_2\rangle,$ and so on. All such single configurations can be collected into the set 
\begin{equation}
\label{setBHF}
\{|N,0\rangle,\ |N-1,1\rangle,\ |N-2,2\rangle,\ \ldots,\ |0,N\rangle\},
\end{equation}
which, like (\ref{set}), spans an $(N+1)$-dimensional Fock space but, in addition, has states that are now specified by BHF equations.

Considering that the eigenstates of the many-body Hamiltonian (\ref{hamtwomode}) are not, in general, a single configuration but are rather composed of a linear combination of $N+1$ such configurations, it is evident that a single configurational description is quite limiting. In particular, it lacks all effects of correlation that arise between atoms in different configurations. To this end, we formulate a new approach in Section IV that combines the full diagonalization of the many-body Hamiltonian in a restricted basis with mean field theory for the underlying single particle states.


\section{Multiconfigurational Bosonic Hartree-Fock Theory}
The many-boson theory restricted to a Fock basis consisting of two single-particle states and the two-single-particle state BHF mean field theory provide complementary descriptions of the BEC and its fragmentation into two condensates. For arbitrary interaction strength, the BHF approach is well justified whenever the state of the BEC can be described by a single configuration, but breaks down whenever a multiconfigurational description is appropriate. Alternatively, the finite basis representation of the many-body Schr\"odinger equation accounts for atomic correlation between each Fock state and includes the effects of the condensate mean field directly in the Hamiltonian. However, it does not provide equations that specify the underlying single-particle wavefunctions. The use of parameters or even single-particle Schr\"odinger wavefunctions may not capture certain properties of the BEC that are associated with strongly interacting atomic gases, where mean field effects on the shape of the single-particle wavefunctions are important.

To this end, we variationally combine the BHF mean field theory of Section III with the restricted Fock state representation of the many-body theory of Section II, allowing for both:
\begin{itemize}
\item[(1.)]
The effects of the condensate mean field on the shape of the single-particle wavefunctions.
\item[(2.)]
The ability to describe states that are made up of multiple configurations.
\end{itemize}
The former is necessary in the strongly interacting regime, while the latter is needed to describe condensate fragmentation within our approach. We refer to the union of these disjoint theories as the {\it multiconfigurational bosonic Hartree-Fock theory} or MCBHF. This theory is rich enough to characterize the atomic structure of the simple BEC and its fragmentation into two condensates at zero temperature as its trapping potential is deformed from a single well to a double well with large barrier height.





\subsection{General Theory}
The basic idea behind our MCBHF approach is to diagonalize a representation of the many-body Hamiltonian (\ref{hamtwomode}) in a set of basis functions of the form
\begin{equation}
\label{gcs}
|N_1,N_2;\{N'_1,N'_2\}\rangle,
\end{equation}
where the total number of atoms $N=N_1+N_2=N'_1+N'_2,$ and $N_1,N'_1=0,\ldots,N.$ These basis states are a combination of the Fock states of Section II and the BHF configurations of Section III. It will become evident that $N_1$ and $N_2$ count atoms in left- and right- localized Fock states, while $N'_1$ and $N'_2$ count atoms in symmetric and antisymmetric BHF states. We call the kets (\ref{gcs}) {\it generalized configuration states} or GCSs because for every underlying BHF configuration $|N'_1,N'_2\rangle,$ which has been indicated in (\ref{gcs}) by $\{N'_1,N'_2\},$ there are an additional $N+1$ Fock states $|N_1,N_2\rangle$ that can be built from this BHF reference configuration. For instance, the subset of GCSs stemming from the particular BHF configuration $|N'_1,N'_2\rangle$ is
\begin{widetext}
\begin{equation}
\label{hfset1}
\{|0,N;\{N'_1,N'_2\}\rangle,\ |1,N-1;\{N'_1,N'_2\}\rangle,\ |2,N-2;\{N'_1,N'_2\}\rangle,\ \ldots,\ |N,0;\{N'_1,N'_2\}\rangle\}.
\end{equation}
However, there are $(N+1)$-many underlying BHF configurations in total. The collection of all GCSs (\ref{gcs}) can be organized into the set
\begin{equation}
\left\{
\begin{array}{lllll}
\label{hfsettotal}
|0,N;\{0,N\}\rangle,&|1,N-1;\{0,N\}\rangle,&|2,N-2;\{0,N\}\rangle,&\ldots,&|N,0;\{0,N\}\rangle,\\
|0,N;\{1,N-1\}\rangle,&|1,N-1;\{1,N-1\}\rangle,&|2,N-2;\{1,N-1\}\rangle,&\ldots,&|N,0;\{1,N-1\}\rangle,\\
|0,N;\{2,N-2\}\rangle,&|1,N-1;\{2,N-2\}\rangle,&|2,N-2;\{2,N-2\}\rangle,&\ldots,&|N,0;\{2,N-2\}\rangle,\\
\vdots&\vdots&\vdots&&\vdots\\
|0,N;\{N'_1,N'_2\}\rangle,&|1,N-1;\{N'_1,N'_2\}\rangle,&|2,N-2;\{N'_1,N'_2\}\rangle,&\ldots,&|N,0;\{N'_1,N'_2\}\rangle,\\
\vdots&\vdots&\vdots&&\vdots\\
|0,N;\{N,0\}\rangle,&|1,N-1;\{N,0\}\rangle,&|2,N-2;\{N,0\}\rangle,&\ldots,&|N,0;\{N,0\}\rangle
\end{array}
\right\}
\end{equation}
containing $(N+1)\times(N+1)$-many elements of which (\ref{hfset1}) makes up just one row. These elements span a Fock space that contains the $(N+1)$-dimensional Fock spaces of Sections II and III in certain limits. For example, the set of all GCSs (\ref{hfsettotal}) reduces to the set (\ref{set}) associated with the Schr\"odinger model of Section II in the limit of vanishing atom-atom interaction within the BHF equations. This corresponds to the subset of all $N+1$ columns belonging to any single row of (\ref{hfsettotal}), since all rows within each column are equivalent in the noninteracting limit of BHF. Alternatively, the subset of GCSs of the form $|N_1,N_2;\{N'_1=N_1,N'_2=N_2\}\rangle$ is identical to the set of BHF configurations (\ref{setBHF}) of Section III. The diagonal entries of (\ref{hfsettotal}) realize such a subset.

For each BHF configuration $|N'_1,N'_2\rangle,$ which is determined by solving the BHF equations (\ref{coupledgpe}), the MCBHF state vector can be expanded onto the GCS basis (\ref{gcs}) according to
\begin{equation}
|\Psi^N;\{N'_1,N'_2\}\rangle_{\nu}=\sum_{N_1=0}^{N}C^{\nu}_{N_1}[N'_1,N'_2]|N_1,N-N_1;\{N'_1,N'_2\}\rangle,
\end{equation}
where the expansion coefficient $C^{\nu}_{N_1}[N'_1,N'_2]$ is the probability amplitude for the $\nu$th excited state of the system to be in the GCS $|N_1,N_2;\{N'_1,N'_2\}\rangle.$ With the MCBHF states, the many-body Schr\"odinger equation 
\begin{equation}
\hat{H}|\Psi^N;\{N'_1,N'_2\}\rangle_{\nu}=E^\nu[N'_1,N'_2]|\Psi^N;\{N'_1,N'_2\}\rangle_{\nu}
\end{equation}
is represented in the GCS basis (\ref{gcs}). In a symmetric trapping potential $V_{\textrm{ext}},$ the Hamiltonian matrix elements $h_{kl}=\langle k|h|l\rangle$ and $V_{klmn}=\langle kl|V|mn\rangle$ are constructed out of the left- and right-localized states $(1/\sqrt{2})[|\chi_1\rangle\pm|\chi_2\rangle],$ which are obtained from the symmetric and antisymmetric BHF solutions $|\chi_1\rangle$ and $|\chi_2\rangle$ by unitary transformation. Such a transformation can always be performed within $h_{kl}$ and $V_{klmn}$ since they provide only matrix elements for the many-body Hamiltonian (\ref{hamtwomode}). Consequently, the Fock space occupation numbers $N_1$ and $N_2,$ which are the first two entries of the GCS (\ref{gcs}), refer to the number of atoms in left- and right-localized single-particle states, while the BHF occupations numbers $N'_1$ and $N'_2,$ which appear in brackets within (\ref{gcs}), refer to the number of atoms in symmetric and antisymmetric single-particle states. Full diagonalization in this basis gives rise to a set of energy eigenvalues $E^\nu[N'_1,N'_2]$ and associated eigenvectors $C^{\nu}_{N_1}[N'_1,N'_2]$ that depend upon the particular BHF reference configuration $|N'_1,N'_2\rangle.$ Solving this eigenvalue problem for each of the $N+1$ configurations results in the set of all MCBHF energies
\begin{equation}
\label{Eset1}
\{E^\nu[0,N],\ E^\nu[1,N-1],\ E^\nu[2,N-2],\ \ldots,\ E^\nu[N'_1,N'_2],\ \ldots,\ E^\nu[N,0]\},
\end{equation}
\end{widetext}
where $\nu=0,\ldots,N.$ The cardinality of this set is $(N+1)\times(N+1).$

We then appeal to the variational principle \cite{Merzb} and Hylleraas-Undheim theorem \cite{Newton1966a,Peierls1979a,Hyll30,Mac33}, which is presented in the Appendix without proof. The variational principle asserts that within the model space of MCBHF, the ground state energy $E^0$ stemming from any BHF configuration is an upper bound to the exact ground state energy and may be systematically reduced by adding more GCSs. Furthermore, by the Hylleraas-Undheim theorem, we know that $\nu$th MCBHF excited state energy $E^\nu$ stemming from any BHF configuration is an upper bound to the exact $\nu$th excited state energy, and may also be systematically reduced by adding more GCSs to the basis. Therefore, it is true in general, that the optimal BHF configuration at each energy level is the one that minimizes the energy at that particular level. By choosing the BHF configuration $|N'_1,N'_2\rangle$ that minimizes $E^0,$ and the BHF configuration $|N''_1,N''_2\rangle$ that minimizes $E^1,$ and so on, we generate a set of $N+1$ optimal MCBHF energies 
\begin{equation}
\label{Eset2}
\{E^0[N'_1,N'_2],\ E^1[N''_1,N''_2],\ \ldots,\ E^N[N'''_1,N'''_2]\},
\end{equation}
where each energy $E^\nu$ may stem from a different BHF reference configuration. In fact, for each configuration, the coupling constant $g$ in the BHF equations (\ref{pcoupledgpe}) could also be treated as a variational parameter \footnotemark[\value{footnote}], but in this case $g$ may not be the same as the physical coupling constant. Each of these optimal energies has an associated optimal eigenvector that is based off of the same BHF configuration that optimizes the energy. For example, the set of eigenvectors corresponding to the set (\ref{Eset2}) is
\begin{equation}
\label{Cset1}
\{C^0_{N_1}[N'_1,N'_2],\ C^1_{N_1}[N''_1,N''_2],\ \ldots,\ C^N_{N_1}[N'''_1,N'''_2]\},
\end{equation}
where $N_1=0,\ldots,N$ and where each of the individual BHF reference configurations $\{N'_1,N'_2\},$ $\{N''_1,N''_2\},$ and $\{N'''_1,N'''_2\}$ are the same in both (\ref{Eset2}) and (\ref{Cset1}). In this way, the MCBHF theory simultaneously optimizes both the underlying basis functions, {\it i.e.}, the BHF reference configurations, and the expansion coefficients of the GCSs. Therefore, each state vector $|\Psi^N;\{N'_1,N'_2\}\rangle_{\nu}$ is potentially made up of multiple GCSs between which arise correlations, and where the effects of the condensate's mean field are built into the Hamiltonian as well as into the shape of the underlying single-particle wavefunctions.

We call the MCBHF state vector that has the lowest MCBHF energy at the $\nu$th energy level the optimal $\nu$th MCBHF state vector and denote it by
\begin{equation}
|\Psi^N;\{N^{\prime\bullet}_1,N^{\prime\bullet}_2\}\rangle_{\nu}.
\end{equation}
The bullets symbolize that the energy of the $\nu$th state is minimized in the BHF configuration $|N'_1,N'_2\rangle.$ The corresponding set of constituent GCSs forms an optimal $(N+1)$-dimensional subset of (\ref{hfsettotal}) that spans a Fock space supporting lower energies than either of those in Sections II and III. We now provide an algorithmic method by which to realize this subset.



\subsection{Implementation Algorithm}
Implementation of the MCBHF theory begins by solving the BHF equations (\ref{pcoupledgpe}) for a particular trap geometry, say a single well potential, and a particular BHF configuration, say the single configuration \footnote{In practice, we do not use the extreme BHF configurations $|0,N\rangle$ and $|N,0\rangle$ as they are GP states and not BHF. Rather in these two limits we choose $|\varepsilon,N-\varepsilon\rangle$ and $|N-\varepsilon,\varepsilon\rangle,$ where $\varepsilon=0.01.$ Note that solution of the BHF equations does not restrict $N'_1$ and $N'_2$ to be integers.}
\begin{equation}
|N'_1=0,N'_2=N\rangle\equiv|\Psi^{\textrm{BHF}}\rangle=(\hat b_1^\dagger)^{0}(\hat b_2^\dagger)^{N}|{\textrm{vac}}\rangle/\sqrt{0!N!},
\end{equation}
where there are no atoms in the single-particle state $|\chi_1\rangle$ and $N$ atoms in the single-particle state $|\chi_2\rangle.$ We choose $V_{\textrm{ext}}$ to be symmetric so that the solutions $\chi_1$ and $\chi_2$ of the BHF equations are symmetric and antisymmetric. With these single-particle wavefunctions we first calculate the BHF energy (\ref{BHFE}). We then unitarily transform $\chi_1$ and $\chi_2$ to the left- and right-localized functions $(1/\sqrt{2})[\chi_1\pm\chi_2],$ and use them to build the matrix elements $h_{kl}$ and $V_{klmn}$ in the many-body Hamiltonian (\ref{hamtwomode}). With the MCBHF state vector
\begin{equation}
|\Psi^N;\{0,N\}\rangle_{\nu}=\sum_{N_1=0}^{N}C^{\nu}_{N_1}[0,N]|N_1,N-N_1;\{0,N\}\rangle,
\end{equation}
full diagonalization of the Hamiltonian yields a basis representation of the ground and excited state energies and associated eigenvectors stemming from the single BHF configuration $|N'_1=0,N'_2=N\rangle.$

Next, we solve the BHF equations (\ref{pcoupledgpe}) associated with the single configuration $|N'_1=1,N'_2=N-1\rangle=(\hat b_1^\dagger)^{1}(\hat b_2^\dagger)^{N-1}|{\textrm{vac}}\rangle/\sqrt{1!(N-1)!},$ in which a single atom has been moved from the second single-particle state to the first. With the BHF solution for this configuration we again calculate $E^{\textrm{BHF}},$ make a unitary transformation to left- and right-localized states, and then build the matrix elements $h_{kl}$ and $ V_{klmn}.$  With the MCBHF state vector
\begin{equation}
\begin{split}
|\Psi^N;&\{1,N-1\}\rangle_{\nu}\\
&=\sum_{N_1=0}^{N}C^{\nu}_{N_1}[1,N-1]|N_1,N-N_1;\{1,N-1\}\rangle
\end{split}
\end{equation}
the full diagonalization of the associated Hamiltonian yields the MCBHF ground and excited state energies and eigenvectors stemming from $|N'_1=1,N'_2=N-1\rangle.$

\begin{figure}
\psfrag{N1/N}[][]{{\large $N'_1/N$}}
\psfrag{CI and MF energy}[][]{{\large MCBHF energy ($\hbar\omega$)}}
\rotatebox{0}{\resizebox{!}{6cm}{\includegraphics{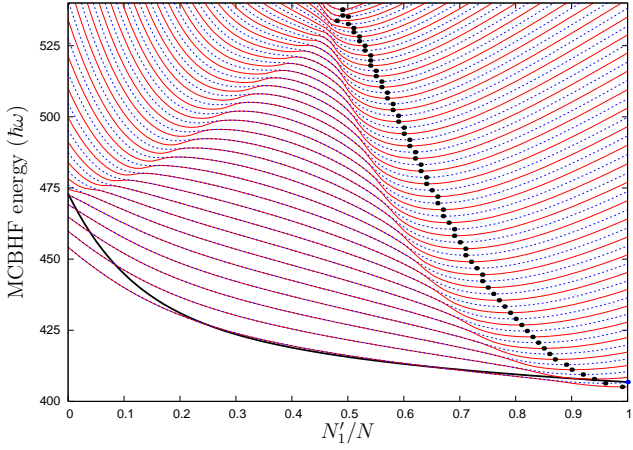}}}
\caption{\label{zoom}(Color online) MCBHF ground and excited state energies and BHF ground state energy for $N=100$ atoms in a single well potential (or zero barrier limit of a double well) with a dimensionless coupling of $\alpha_{\textrm{Q1D}}=40$ [see (\ref{gq1d})] $\alpha_{\textrm{Q1D}}=40$ [see (\ref{gq1d})] versus BHF reference configuration. The solid black curve represents the BHF ground state energy. Bullets are placed at the minima of BHF ground and MCBHF ground and excited energies. The line style is alternated to aid in visualization. This figure is a zoomed in and more detailed version of the first panel in Figure \ref{E_vs_N}.}
\end{figure}

We then repeat this procedure for each BHF configuration until we reach the last BHF configuration $|N,0\rangle=(\hat b_1^\dagger)^{N}(\hat b_2^\dagger)^{0}|{\textrm{vac}}\rangle/\sqrt{N!0!}$ in which all $N$ atoms are in the symmetric state $|\chi_1\rangle$ and no atoms are in the antisymmetric state $|\chi_2\rangle.$ Mapping out the complete set of MCBHF energies versus all possible BHF reference configurations at this trap geometry results in a diagram like that displayed in Figure \ref{zoom}.
Notice the detailed energy level mergings and splittings that occur in this figure. By appealing to the variational and Hylleraas-Undheim theorems, we find the variationally optimal MCBHF ground state and excited state energies and denote their location by a single bullet. The location of these bullets also indicates which BHF configuration provides the optimal MCBHF ground and excited state energy. In this way, we are variationally optimizing both the shape of the single-particle wavefunctions and the expansion coefficients. That is, a nonlinear optimization determines the shape of the BHF single-particle wavefunctions for each BHF configuration while a second linear Hylleraas-Undheim optimization determines the expansion coefficients of the optimal MCBHF state vector $|\Psi^{N};\{N^{\prime\bullet}_1,N^{\prime\bullet}_2\}\rangle_{\nu}.$ We then repeat this algorithm as the trapping potential is deformed from a single to a double well geometry.

\subsection{Example: MCBHF Excited State}
\begin{figure}
\psfrag{N1/N}[][]{{\large $N_1/N$}}
\psfrag{CI eigenvectors}[][]{{\large $C^{51}_{N_1}[56,44]$ expansion coefficients}}
\rotatebox{0}{\resizebox{!}{6cm}{\includegraphics{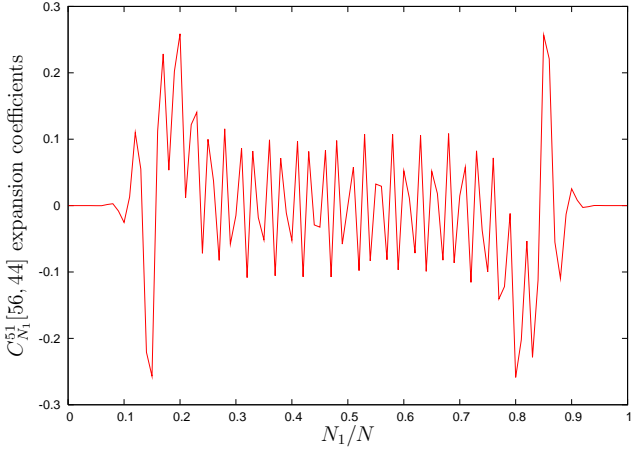}}}
\caption{\label{cnfig}(Color online) MCBHF expansion coefficients $C^{51}_{N_1}$ of the 51st excited state $|\Psi^{N};\{56^\bullet,44^\bullet\}\rangle_{51}$ versus the occupation number in each GCS. The label $\{56,44\}$ in the state vector signifies that the minimal 51st MCBHF excited state energy has been attained in the BHF configuration $|N'_1=56,N'_2=44\rangle.$ This probability amplitude corresponds to 100 atoms in a single well trapping potential with $\alpha_{\textrm{Q1D}}=40$ [see (\ref{gq1d})].}
\end{figure}

Consider, for example, the optimal 51st excited state in Figure \ref{zoom}. It is built off of the $|N'_1=56,N'_2=44\rangle$ BHF reference configuration, which can be read off of Figure \ref{zoom}, and has expansion coefficients $C^{51}_{N_1}[56,44]$ as shown in Figure \ref{cnfig}. 
\begin{figure}
\psfrag{psi}[][]{{\normalsize $\chi_1$ and $\chi_2$ $(\beta^{-1/2})$}}
\psfrag{psi1}[][]{\rotatebox{180}{{\normalsize $V_{\textrm{ext}}$ $(\hbar\omega)$}}}
\psfrag{x}[][]{{\normalsize$z$ $(\beta)$}}
\psfrag{mode 1}[][]{$\chi_1$\hspace{-0.3cm}}
\psfrag{mode 2}[][]{$\chi_2$\hspace{-0.3cm}}
\psfrag{TF}[][]{TF\hspace{0.3cm}}
\psfrag{Vex}[][]{$V_{\textrm{ext}}\hspace{0.3cm}$}
\rotatebox{0}{\resizebox{!}{6cm}{\includegraphics{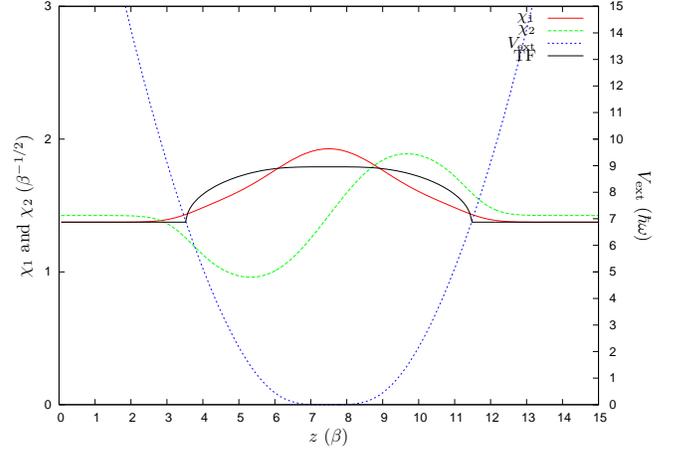}}}
\caption{\label{wf_51}(Color online) BHF single-particle wavefunctions $\chi_1$ and $\chi_2$ versus coordinate $z$ associated with the BHF configuration $|N'_1=56,N'_2=44\rangle,$ which underlies the optimal 51st MCBHF excited state at a barrier height of 0 $\hbar\omega.$ A dimensionless coupling of $\alpha_{\textrm{Q1D}}=40$ [see (\ref{gq1d})] for 100 atoms was used. Both wavefunctions have been set at the chemical potentials $\mu_{11}$ and $\mu_{22}$ associated with this configuration. The corresponding Thomas-Fermi wavefunction is plotted as a solid black curve.}
\end{figure}
This particular optimal MCBHF excited state is not described by the single BHF configuration $|N'_1=56,N'_2=44\rangle$ displayed in Figure \ref{wf_51} but rather by almost all possible GCSs within this model space. It is roughly of the form
\begin{widetext}
\begin{equation}
|\Psi^{N};\{56^\bullet,44^\bullet\}\rangle_{51}\approx C^{51}_{10}|10,90;\{56,44\}\rangle+C^{51}_{11}|11,89;\{56,44\}\rangle+\cdots+C^{51}_{56}|56,44;\{56,44\}\rangle+\cdots+C^{51}_{90}|90,10;\{56,44\}\rangle,
\end{equation}
\end{widetext}
where $C^{51}_{N_1}\equiv C^{51}_{N_1}[56,44]$ and where the label $\{56,44\}$ in each ket signifies that the matrix elements of the many-body Hamiltonian (\ref{hamtwomode}) for the optimal 51st excited state have been built out of the single BHF configuration $|N'_1=56,N'_2=44\rangle,$ where $N=100=N_1+N_2.$ While this BHF configuration minimizes the energy for the 51st excited state, another BHF configuration may be optimal for the 52nd, and so on. Furthermore, each optimal state may also have a different distribution of expansion coefficients $C^\nu_{N_1}.$ In other words, every optimized MCBHF state is potentially derived from a different BHF reference configuration and is potentially made up of a unique superposition of GCSs.

\section{Results and Discussion}

We employ an external trapping potential $V_{\textrm{ext}}$ that is harmonic with a Gaussian function centered at the oscillator's minimum. In quasi-one dimension \cite{Carr00,Olshanii1998a}, it has the dimensionless form
\begin{equation}
\label{vext}
V_{\textrm{ext}}(z)=(1/2)[z-\ell_z/2]^2+A\exp(-[z-\ell_z/2]^2/2),
\end{equation}
where $\ell_z$ is the spatial grid length and $A$ is the amplitude of the Gaussian having unit width in oscillator units. While we do not intend to precisely model a particular experiment in this paper, the functional form of $V_{\textrm{ext}}$ \cite{Salasnich1999a} has been chosen because of its resemblance to the double well interference experiments performed at MIT \cite{Shin2004a,Saba05,Andrews1997b}. In practice, $A=1+\varepsilon B$ and $B$ is varied from zero up to 150 in increments of $\varepsilon=0.1$ $\hbar\omega.$ As the amplitude is increased, the constant $1+\ln A$ is subtracted off of the potential so that its minimum always lies at 0 $\hbar\omega.$ All energies are, therefore, reported as relative to the zero of the trapping potential. In this way, $V_{\textrm{ext}}$ is deformed from an initially single well with zero barrier height to a double well with a barrier height of $\varepsilon B=15$ $\hbar\omega.$ 


In the GP equation (\ref{gpe1}), varying $g$ is the same as varying $N.$ However, $g$ and $N$ enter the many-body Hamiltonian (\ref{hamtwomode}) in different ways so that varying $g$ is not the same as varying $N$ in MCBHF theory. Although actual experiments involve $N\approx10^5$ condensate atoms, we choose to perform all calculations with a smaller number. Nevertheless, it is possible to illustrate the importance of correlation and mean field effects by adjusting $g$ so that the product $gN$ has the correct value. In quasi-one-dimension, the dimensionless coupling constant becomes
\begin{equation}
\label{gq1d}
\alpha_{\textrm{Q1D}}\equiv4\pi a\beta_zN/L^2_\perp,
\end{equation}
where $\beta_z=\sqrt{\hbar/m\omega_z}$ is the $z$-oscillator length and $L_\perp$ is the transverse length associated with the quasi-one-dimensional approximation. Choosing trap frequency $\omega_z=2\pi\times30$ Hz and $S$-wave scattering length $a=4.9$ nm appropriate for the recent ${^{23}\textrm{Na}}$ double well interference experiments \cite{Shin2004a,Saba05}, the dimensionless quasi-one-dimensional coupling constant has an approximate value of 50, where a condensate density of $10^{13}$ $\textrm{atoms}/(\textrm{cm})^3$ has been assumed. In light of this value, we perform MCBHF calculations at $\alpha_{\textrm{Q1D}}=40$ where Thomas-Fermi characteristics are already evident. The quasi-one-dimensional BHF wavefunctions displayed in Figure \ref{hfwffig} correspond to this particular value of coupling strength. Throughout the paper, we have chosen the frequency scale $\omega$ to be $\omega_z.$


\subsection{MCBHF energy versus BHF configuration}
\begin{figure*}
\psfrag{N1/N}[][]{{\large $N'_1/N$}}
\psfrag{CI and MF energy}[][]{{\large MCBHF energy ($\hbar\omega$)}}
\rotatebox{0}{\resizebox{!}{12.9cm}{\includegraphics{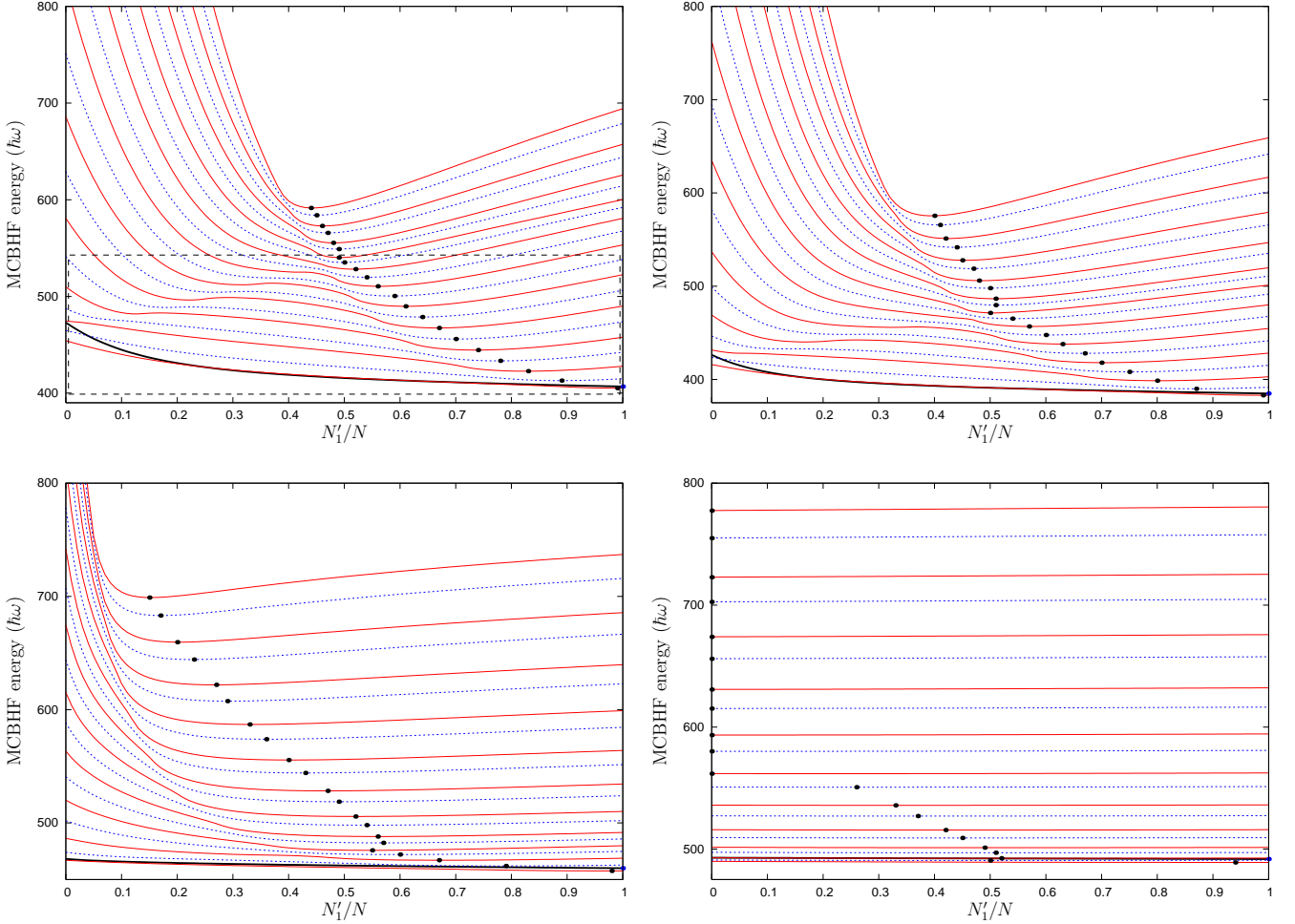}}}
\caption{\label{E_vs_N}(Color online) MCBHF ground and excited state energies and BHF ground state energy for $\alpha_{\textrm{Q1D}}=40$ and $N=100$ versus BHF configuration. Four different barrier heights 0, 3, 9, and 13 $\hbar\omega$ are displayed from left to right and top to bottom. The solid black curve in each panel represents the BHF ground state energy at each barrier height and for each BHF configuration. Bullets are placed at the minima of the BHF ground state and all MCBHF energies. These are the variationally optimal states at each energy level. To aid in visualization, we alternate line style and only display every fifth excited state. The box in the first panel corresponds to Figure \ref{zoom}.}
\end{figure*}

Allowing the interwell barrier in (\ref{vext}) to grow, we repeat the implementation algorithm of Section IV B for each and every barrier height. A panorama of MCBHF energies versus BHF configuration is presented in Figure \ref{E_vs_N} corresponding to four different barrier heights 0, 3, 9, and 13 $\hbar\omega$ with $\alpha_{\textrm{Q1D}}=40$ in both the many-body Hamiltonian (\ref{hamtwomode}) and in the BHF equations (\ref{pcoupledgpe}). The first panel is an enlarged version of Figure \ref{zoom}, where only every fifth line is plotted to aid in visualization.
Bullets have again been placed at each of the variationally optimal energies. The structure of the MCBHF ground and excited state energies versus BHF reference configuration is quite detailed and it is important to explain some of its features. For example, it can be seen in Figure \ref{zoom} or in the upper two panels of Figure \ref{E_vs_N} that there are two energetic pathways that can support optimal solutions. One pathway provides the globally optimal solutions while the second pathway provides only locally optimal solutions. One might ask why this structure is present and why it is not symmetric about the $|N/2,N/2\rangle$ BHF configuration. The answer lies in the role of the two single-particle states $|\chi_1\rangle$ and $|\chi_2\rangle,$ which are the symmetric and antisymmetric BHF solutions, as various numbers of atoms are placed in each state.

Consider the BHF reference configuration $|0,N\rangle$ in which there are no atoms in the symmetric state and $N$ atoms in the antisymmetric state. Excited MCBHF states $|\Psi^N;\{0,N\}\rangle_\nu$ are made up of single atom excitations out of the highest energy BHF reference configuration describable in the basis. Furthermore, these excitations move atoms to states that have zero occupation. Due to the effect of bosonic amplification, these excitations of an excited state cost a macroscopic amount of energy \cite{Huang1957a}. Alternatively, consider the MCBHF state $|\Psi^N;\{N,0\}\rangle_\nu,$ which is built out of the BHF reference configuration $|N,0\rangle,$ in which all atoms are in the symmetric ground state. Single atom excitations out of this state also cost a macroscopic amount of energy, due to bosonic amplification, as atoms are moved from a macroscopically filled state to empty states. However, these excitation energies lie lower than the previous energies because they correspond to excitations out of the ground state rather than excitations from an excited state. This explains why the energies on the left side of each panel, {\it i.e.}, $N'_1/N\sim0,$ are larger than those on the right, {\it i.e.}, $N'_1/N\sim1.$ Lastly, consider the MCBHF state $|\Psi^N;\{N'_1\approx N/2,N'_2\approx N/2\}\rangle_\nu,$ which is approximately built out of the BHF reference configuration $|N/2,N/2\rangle,$ in which both states are macroscopically occupied. Single atom excitations out of $|\Psi^N;\{N/2,N/2\}\rangle_\nu$ {\it exchange} atoms between two macroscopically occupied states. In distinction to the previous two cases, here there is a macroscopic energy cost to move a single atom out of $|\chi_1\rangle$, but there is also a macroscopic energy gain as the atom is added to $|\chi_2\rangle.$ Therefore, the MCBHF energies in this region represent the smallest energy excitations possible within our model. The location of each minimum is biased to right because it is energetically easier to make excitations out of the ground state than out of excited states. This is why the energy pathway on the right provides the global minimum solutions while the energy pathway on the left supports only local minima.

\subsection{Optimal MCBHF energy versus barrier height}
\begin{figure*}
\psfrag{MCSCF energy}[][]{{\normalsize MCBHF energy $(\hbar\omega)$}}
\psfrag{barrier height}[][]{{\normalsize barrier height $(\hbar\omega)$}}
\rotatebox{0}{\resizebox{!}{6cm}{\includegraphics{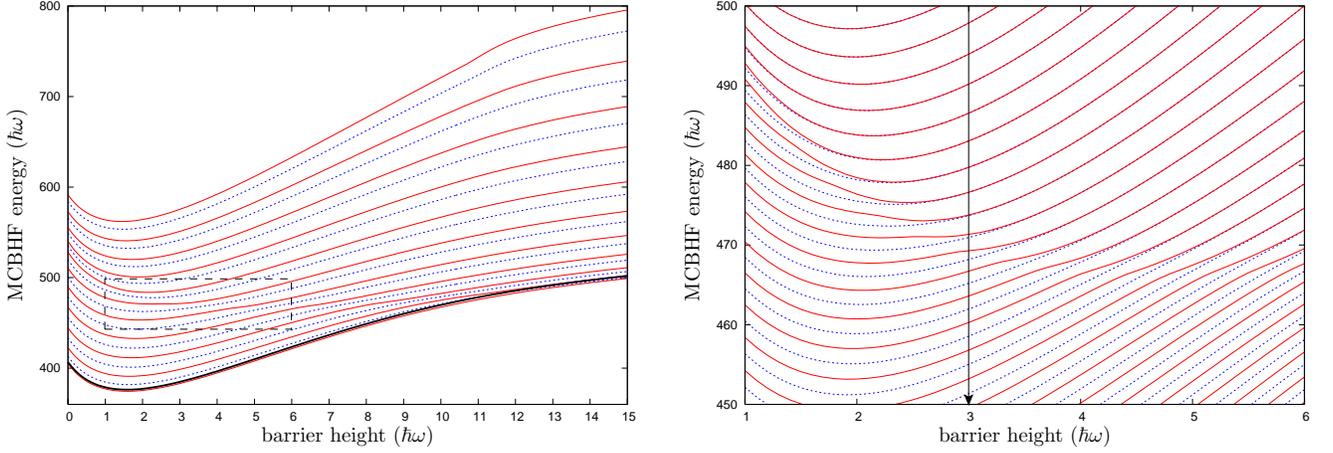}}}
\caption{\label{corrhf}(Color online) MCBHF energy level correlation diagram versus barrier height for $\alpha_{\textrm{Q1D}}=40$ and $N=100$ atoms. Every fifth energy level is plotted in the left panel and the line styles are alternated in both to aid in visualization. The solid black curve in the left panel denotes the BHF ground state energy. Energy level mergings, which lead to a pronounced ridge structure separating nondegenerate oscillator like states from doubly degenerate macroscopic superposition states, are focused on in the second panel. This panel is a zoomed and more detailed version of the boxed region in the first, where every line has been displayed.}
\end{figure*}

By collecting the set of all optimal energies at each barrier height, {\it i.e.}, by collecting all the bulleted energies appearing in Figure \ref{E_vs_N} plus those at all other barrier heights, an energy level correlation diagram similar to those in Figures \ref{heidiplot} and \ref{sch} can be made. The MCBHF correlation diagram associated with Figure \ref{E_vs_N} is displayed in Figure \ref{corrhf}.
\begin{figure*}
\psfrag{MCSCF energy}[][]{{\normalsize MCBHF energy $(\hbar\omega)$}}
\psfrag{barrier height}[][]{{\normalsize barrier height $(\hbar\omega)$}}
\rotatebox{0}{\resizebox{!}{6cm}{\includegraphics{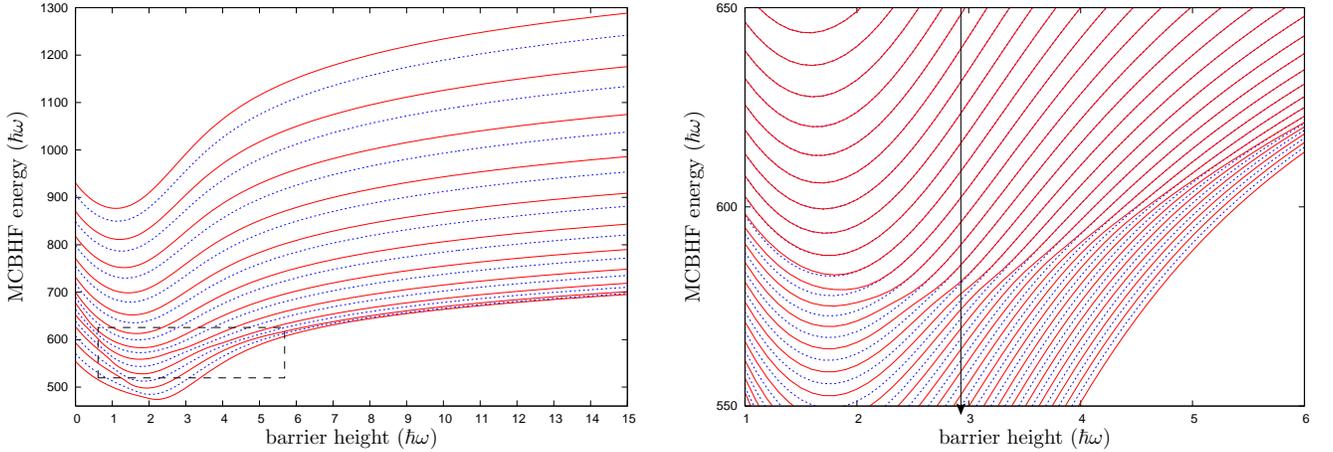}}}
\caption{\label{corrhf0.1}(Color online) Schr\"odinger energy level correlation diagram versus barrier height for $\alpha_{\textrm{Q1D}}=40$ and $N=100$ atoms. Every fifth energy level is plotted in the left panel and the line styles are alternated in both to aid in visualization. The second panel is a zoomed and more detailed version of the boxed region in the first, where every line has been displayed.}
\end{figure*}
There are qualitative similarities to those of the parameterized and Schr\"odinger model calculations such as, for example, the pronounced ridge structure that marks the merging of nondegenerate energy levels to degenerate as the interwell barrier is raised. However, large quantitative differences are apparent by comparing the MCBHF correlation diagram in Figure \ref{corrhf} with the corresponding Schr\"odinger correlation diagram displayed in Figure \ref{corrhf0.1}. The latter figures are computed by diagonalizing the many-body Hamiltonian (\ref{hamtwomode}) with $\alpha_{\textrm{Q1D}}=40,$ where the underlying single-particle wavefunctions are taken from the single-particle Schr\"odinger equation. No optimization is necessary in this case since the wavefunctions have no dependence upon $N'_1$ and $N'_2.$ To aid in visualization, the second panels in Figures \ref{corrhf} and \ref{corrhf0.1} have been zoomed in precisely at the phase transition between nondegenerate and degenerate energies. Notice the large difference in energy scales as well as the shape and placement of the phase transitions in Figures \ref{corrhf} and \ref{corrhf0.1}.

\subsection{Distribution of GCSs and BHF wavefunctions}
\begin{figure*}
\psfrag{gs}[][]{gs}
\psfrag{1es}[][]{1es}
\psfrag{2es}[][]{2es}
\psfrag{63es}[][]{63es}
\psfrag{64es}[][]{64es}
\psfrag{65es}[][]{65es}
\psfrag{CI eigenvectors}[][]{{\large $C^{\nu}_{N_1}$ expansion coefficients}}
\psfrag{N1/N}[][]{{\large $N_1/N$}}
\rotatebox{0}{\resizebox{!}{12.9cm}{\includegraphics{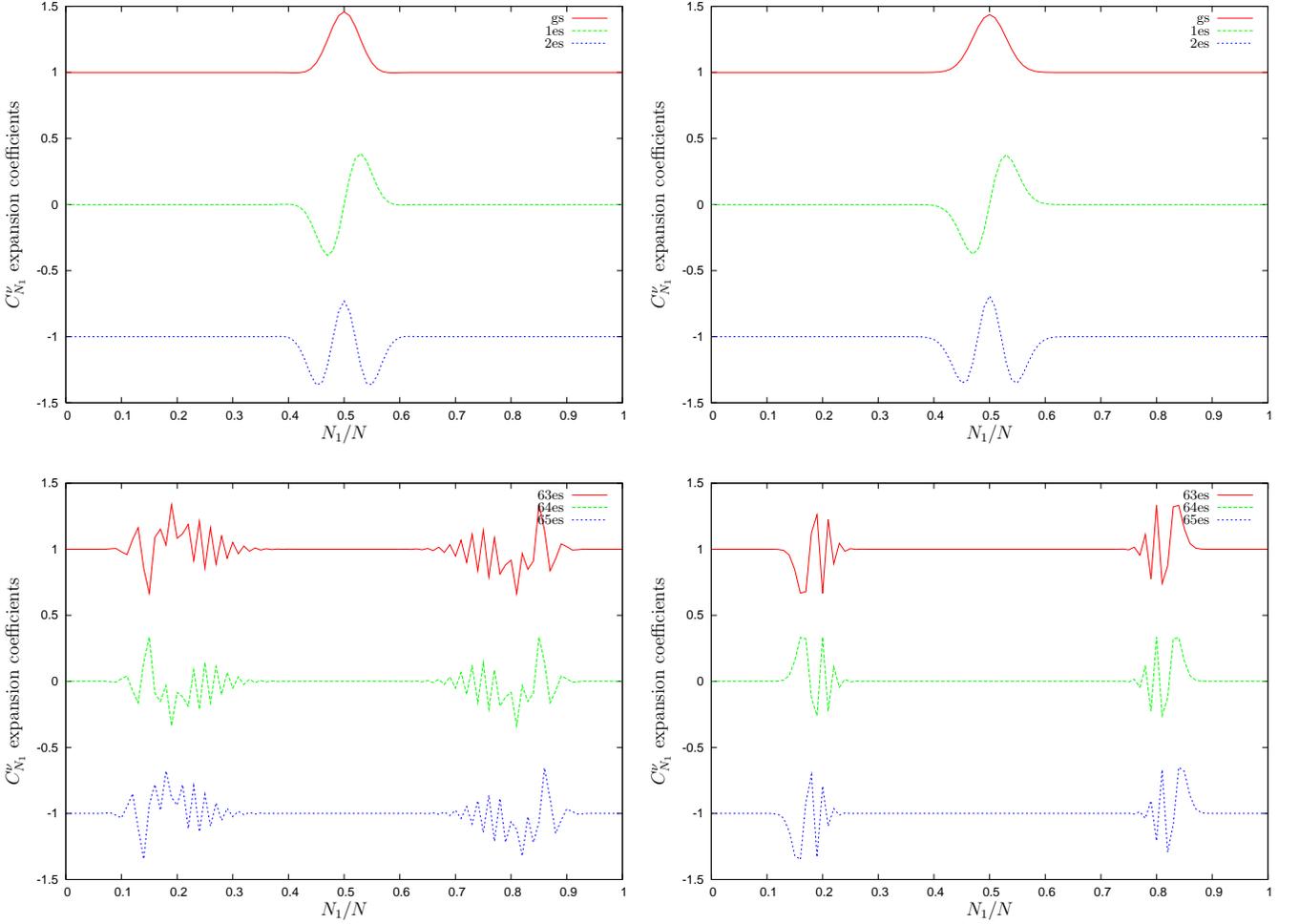}}}
\caption{\label{cn_compare}(Color online) Left two panels: Expansion coefficients of the optimal MCBHF ground state and first and second excited states versus occupation number in each GCS as well as expansion coefficients of the optimal 63rd, 64th, and 65th excited states versus occupation number in each GCS. The optimal BHF reference configurations $|N'_1,N'_2\rangle$ have been provided for each state separately within the text. Right two panels: Expansion coefficients of the Schr\"odinger ground state and first and second excited states versus occupation number as well as expansion coefficients of the Schr\"odinger 63rd, 64th, and 65th excited states versus occupation number. These probability amplitudes correspond to $N=100$ atoms at a barrier height of 3 $\hbar\omega.$ The left column has the dimensionless coupling $\alpha_{\textrm{Q1D}}=40$ in both the many-body Hamiltonian and BHF equations, while the right column has $\alpha_{\textrm{Q1D}}=40$ only in the many-body Hamiltonian.}
\end{figure*}

Examining the distribution of GCSs in each of the optimal MCBHF states in Figure \ref{corrhf}, we find that below the ridge the distribution is harmonic oscillator like in form, while exotic macroscopic superpositions of GCSs exist above the ridge. This behavior is generic across all barrier heights where ridge structures are present. In order to exemplify these characteristics, we present optimal MCBHF expansion coefficients in Figure \ref{cn_compare} taken from Figure \ref{corrhf} at the particular barrier height of 3 $\hbar\omega.$ 
Coefficients are presented that correspond to the optimal MCBHF ground state and first and second excited states, which are below the ridge, as well as to the optimal MCBHF 63rd, 64th, and 65th excited states, which are above the ridge. An analogous presentation is made in Figure \ref{cn_compare} for the Schr\"odinger based coefficients associated with Figure \ref{corrhf0.1}. The two left panels in Figure \ref{cn_compare} correspond to the full MCBHF theory with an interaction strength of $\alpha_{\textrm{Q1D}}=40,$ while the two right panels correspond to the case where single-particle Schr\"odinger wavefunctions are used to build the many-body Hamiltonian with $\alpha_{\textrm{Q1D}}=40.$

Below the ridge, the optimal MCBHF states are not energetically degenerate. The ground state and first and second excited states are optimized in the BHF reference configurations
\begin{equation}
\label{config1}
\begin{array}{lll}
|N'_1=99,N'_2=1\rangle&&\textrm{ground state}\\
|N'_1=95,N'_2=5\rangle&&\textrm{first excited state}\\
|N'_1=93,N'_2=7\rangle&&\textrm{second excited state}.
\end{array}
\end{equation}
Reference configurations do not enter the Schr\"odinger based theory as the single-particle wavefunctions have no dependence on $N'_1$ and $N'_2.$ It is apparent that the ground state is described by a binomial distribution of states peaked around $N_1=50$ and $N_2=50$ in the basis where $N_1$ and $N_2$ are the occupation numbers of states that are left- and right-localized in the trapping potential. However, by making a unitary transformation directly on the coefficients of the GCSs taking them back to the symmetric and antisymmetric basis, the expansion coefficients are tightly peaked around a single state. The optimal MCBHF ground state becomes
\begin{equation}
\label{catstate0}
|\Psi^{N};\{99^\bullet,1^\bullet\}\rangle_0\approx|100,0;\{99,1\}\rangle
\end{equation}
while a similar result holds for the Schr\"odinger case.

\begin{figure*}
\psfrag{psi}[][]{{\large $\chi_1$ and $\chi_2$ $(\beta^{-1/2})$}}
\psfrag{psi1}[][]{\rotatebox{180}{{\large $V_{\textrm{ext}}$ $(\hbar\omega)$}}}
\psfrag{x}[][]{{\large $z$ $(\beta)$}}
\psfrag{mode 1}[][]{$\chi_1$\hspace{-0.3cm}}
\psfrag{mode 2}[][]{$\chi_2$\hspace{-0.3cm}}
\psfrag{TF}[][]{TF\hspace{0.3cm}}
\psfrag{Vex}[][]{$V_{\textrm{ext}}\hspace{0.3cm}$}
\rotatebox{0}{\resizebox{!}{6.25cm}{\includegraphics{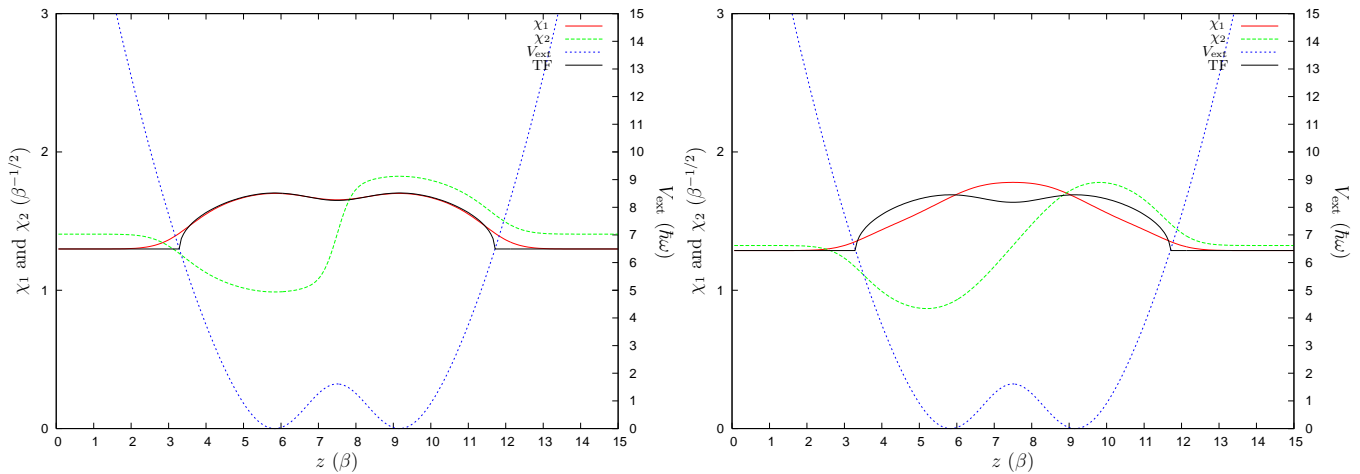}}}
\caption{\label{Re_compare}(Color online) BHF single-particle wavefunctions $\chi_1$ and $\chi_2$ versus coordinate $z$ at a barrier height of 3 $\hbar\omega.$ The first panel corresponds to the BHF configuration that optimizes the MCBHF ground state. That is, the first panel depicts the BHF configuration $|99,1\rangle$ associated with $\alpha_{\textrm{Q1D}}=40.$ The second panel correspond to the BHF configuration that optimizes the 63rd MCBHF excited state. That is, the second panel depicts the BHF configuration $|50,50\rangle$ associated with $\alpha_{\textrm{Q1D}}=40.$ All wavefunctions have been set at the chemical potentials $\mu_{11}$ and $\mu_{22}$ associated with each configuration. At each barrier height, we plot the corresponding Thomas-Fermi wavefunction as a solid black curve.}
\end{figure*}

The underlying BHF single-particle wavefunctions associated with (\ref{catstate0}) are displayed in the first panel of Figure \ref{Re_compare}.
It is important to note that these BHF wavefunctions depend strongly on the number of atoms in each single-particle state as well as on the interaction strength and do look quite different from the corresponding solutions of the single-particle Schr\"odinger equation in the second panel of Figure \ref{hfwffig_sch}. Being that the MCBHF states (\ref{catstate0}) are made up of a single configuration with almost all $N=100$ atoms in $|\chi_1\rangle,$ it is not surprising that GP theory \cite{Gross1961a,Pitaevskii1961a} provides a good description of the ground state of the simple BEC.


A harmonic oscillator like distribution of expansion coefficients continues from the ground state up to and including the excited states that lie within the ridge at approximately 470 $\hbar\omega$ in Figure \ref{corrhf}. However, above the ridge, the distribution of GCSs making up each optimal MCBHF state takes on a striking new form. Macroscopic quantum superpositions states emerge in the spectrum. These states, which are pairwise degenerate, are also known as entangled number states and are colloquially called Schr\"odinger cats. The lower left panel in Figure \ref{cn_compare} depicts the distribution of GCSs in the optimal 63rd, 64th, and 65th excited MCBHF states at a barrier height of 3 $\hbar\omega$ corresponding to $\alpha_{\textrm{Q1D}}=40.$ The lower right panel depicts the associated distribution of Schr\"odinger Fock states. It is found that the former three states are optimized in the BHF reference configurations 
\begin{equation}
\label{config3}
\begin{array}{lll}
|N'_1=50,N'_2=50\rangle&&\textrm{63rd excited state}\\
|N'_1=50,N'_2=50\rangle&&\textrm{64th excited state}\\
|N'_1=50,N'_2=50\rangle&&\textrm{65th excited state}.
\end{array}
\end{equation}
The underlying BHF single-particle wavefunctions for the 63rd excited state are displayed in the second panel of Figure \ref{Re_compare}. Once again, it is important to note that these BHF wavefunctions depend strongly on the number of atoms in each single-particle state as well as on the interaction strength and do look quite different from the associated solutions of the single-particle Schr\"odinger equation displayed in the second panel of Figure \ref{hfwffig_sch}. Note that the Schr\"odinger based ground and all excited states stem from the same Schr\"odinger wavefunction. The optimal MCBHF excited states are of the approximate form
\begin{equation}
\label{catstate}
\begin{split}
|\Psi^{N};\{50^\bullet,50^\bullet\}\rangle_{\nu}&\approx f^\nu_{15}|15,85;\{50,50\}\rangle\\
&\ \ \ \pm g^\nu_{85}|85,15;\{50,50\}\rangle\\
\end{split}
\end{equation}
where $\nu=63,64,65.$ The coefficients $f^\nu_{15}$ and $g^\nu_{85}$ are meant to indicate that these states are not made up of just two GCSs, but rather there is a distribution of GCSs peaked around $|15,85;\{50,50\}\rangle$ and $|85,15;\{50,50\}\rangle$ as is evident from the lower left panel of Figure \ref{cn_compare}. These states are, therefore, multiconfigurational in nature. In Figure \ref{cn_compare}, the ``+'' combination is for $\nu=64$ and the ``$-$'' combination is for $\nu=63,65.$ Sharper and more extreme macroscopic superposition states do appear in the spectrum at even higher lying energies. Notice that many more configurations are involved in the lower left panel of Figure \ref{cn_compare} than Fock states in the lower right.

In addition, we point out that in the large barrier height limit, the BHF reference configuration underlying the optimal MCBHF ground state and low lying excited states approaches $|N'_1=50,N'_2=50\rangle.$ This can be seen by comparing trends in the lower two panels of Figure \ref{E_vs_N}, which correspond to barrier heights of 9 $\hbar\omega$ and 13 $\hbar\omega.$ The distribution of GCSs at these barrier heights is also sharply peaked at $N_1=50$ and $N_2=50.$ Therefore, the optimal MCBHF ground state takes the form
\begin{equation}
\label{catstate9}
|\Psi^{N};\{50^\bullet,50^\bullet\}\rangle_0\approx|50,50;\{50,50\}\rangle
\end{equation}
in the large barrier height limit, which is a fragmented BEC state. Since the optimal BHF reference configuration has approximately 50 atoms in each of two single-particles states, it is not surprising that a two-single-particle state mean field theory provides a realistic description of the fragmented BEC ground state \cite{Cederbaum2003a,Cederb04}. The MCBHF state (\ref{catstate9}) is, after all, a single configuration with about $N/2$ atoms in each of two single-particle states.

We have demonstrated that the optimal MCBHF ground states at both zero and large barrier height are described by a single GCS. The low barrier limit is characterized by only one single-particle state. This is why ground state zero temperature properties of the simple BEC are well understood with GP theory alone. The large barrier height limit requires, at minimum, two single-particle states. GP theory, being a one-single-particle state theory fails in this case, however, the two-single-particle state BHF theory provides a good description here. Furthermore, it has been shown that states do exist at all barrier heights that can only be described by superpositions of multiple configurations into which the effects of atom-atom interactions are incorporated. Comparison with the Schr\"odinger model of Section II demonstrates a marked difference in the atomic structure of the condensate due to the neglect of mean field effects on the underlying single-particle wavefunctions. Therefore, the full effects of the mean field as well as correlation must be included in order to gain understanding of the entire process of BEC fragmentation and not just an understanding of the simple or fragmented ground state.


\section{Conclusion}
Multiconfigurational Hartree-Fock theory has been formulated for the many-body problem associated with a gas of identical bosonic atoms trapped at zero temperature in potentials that can be continuously deformed from single to double well geometries. A didactic survey of our approach has been presented which clarifies many of the principles and approximations that are found in other relevant approaches from the literature. In the extreme limit of a single configuration, MCBHF theory recovers two-single-particle state mean field theory \cite{Cederbaum2003a,Cederb04}, which includes the effect of the condensate mean field on the single-particle states but lacks the correlation that arises between configurations. In the opposite extreme limit, our approach recovers the exact diagonalization of the many-body Hamiltonian in a restricted two-state basis of Fock states \cite{Spekkens1999a}. Atomic correlation arises automatically in this case, but the full effects of atom-atom interaction are missing. In particular, the underlying single-particle states have no dependence upon the mean field of the condensate. By incorporating both approaches, the MCBHF theory is capable of describing general many-body states of the system that are made up of a superposition of many configurations into which mean field effects are included.

MCBHF theory has been implemented in a systematic study of the many-body atomic structure of the BEC and its fragmentation in a double well trapping potential. Nonlinear and linear optimization has been utilized in conjunction with the variational and Hylleraas-Undheim theorems to find the optimal ground and excited states of the condensate throughout its break up. A variety of interesting delocalized and localized, single configurational and multiconfigurational states have been found to arise throughout the spectrum at all barrier heights. Contributions from the condensate mean field on the underlying single-particle states as well as correlation effects, which arise between configurations, have been emphasized to be essential in order to obtain a more complete microscopic understanding of the condensate's atomic structure throughout the fragmentation process. Future work will be devoted to the construction of an associated dynamical theory that incorporates the rich atomic structure of the MCBHF formalism. It is intended to apply such a dynamical theory to the accurate modeling of the recent experiments \cite{Shin2004a,Saba05,Anker04,Albiez04} in which the observables are time-dependent condensate densities, rather than stationary state energy levels. However, the importance of mean field effects illustrated here for the energy levels of the many-body system as a function of barrier height clearly indicate that this same mixing of mean field and correlation effects will be important there.

\begin{acknowledgments}
This work was partially funded by the NSF grant PHY-0140091 (W.P.R., P.I.) and by a Dissertation Fellowship from the American Association of University Women (S.B.M.). Furthermore, D.M. would like to acknowledge helpful discussions with Prof. B.D. Esry and Prof. C.H. Greene as well as other participants at the Few- and Many-Body Physics in Quantum Liquids and Gases Conference held at the University of Washington's Institute for Nuclear Theory, regarding the contact potential replacement in BHF theory.
\end{acknowledgments}

\appendix*
\section{Hylleraas-Undheim Theorem and Optimization of Bounds for Excited State Eigenvalues}
A nonrigorous statement of the Hylleraas-Undheim theorem \cite{Hyll30} is as follows: The $N$ energy eigenvalues $E_\nu$ of the Hamiltonian $H_{kl}=\langle g_k|H|g_l\rangle$ represented in the orthonormal basis $\{g_1,\ldots,g_N\}$ can be ordered so that
\begin{equation}
E_1\leq E_2\leq\cdots\leq E_N.
\end{equation}
Each $E_\nu$ is an upper bound to the exact $\nu$th eigenvalue $E_\nu^{\textrm{ex}}$ of the same symmetry, {\it i.e.}, 
\begin{equation}
E_1^{\textrm{ex}}\leq E_1, E_2^{\textrm{ex}}\leq E_2,\cdots, E_N^{\textrm{ex}}\leq E_N.
\end{equation}
The bounds can be systematically improved as $N$ is increased since the energy eigenvalues necessarily move downward or stay the same. For example, the eigenvalues $E'_\nu$ of the Hamiltonian represented in the orthonormal basis $\{g_1,\ldots,g_N,g_{N+1}\},$ which contains one extra basis function, interlace the eigenvalues $E_\nu$ in such a way that
\begin{equation}
E_1^{\textrm{ex}}\leq E_1'\leq E_1, E_2^{\textrm{ex}}\leq E_2'\leq E_2,\cdots, E_N^{\textrm{ex}}\leq E_N'\leq E_N.
\end{equation}
Continued interlacing occurs as more basis functions are added. In this way, bounds for excited states can be independently optimized. A proof of the theorem appears in \cite{Newton1966a,Mac33}, while the more casual reader is referred to \cite{Peierls1979a}.

\bibliography{jila,thesis}

\end{document}